%% file: WZ-WDP3.tex
\documentclass[11pt,journal,onecolumn,draftclsnofoot]{IEEEtran}
\usepackage{epsfig}
\usepackage{graphicx}
\usepackage{amsmath}
\usepackage{amsfonts}
\usepackage{subfig}
\usepackage{appendix}

\long\def\symbolfootnote[#1]#2{\begingroup%
\def\thefootnote{\fnsymbol{footnote}}\footnote[#1]{#2}\endgroup}

\begin{document}


\newcommand{\comb}[2]{{#1 \choose #2}}
\newcommand{\uzer}{\underline{0}}
\newcommand{\uV}{\underline{V}}
\newcommand{\uA}{\underline{A}}
\newcommand{\uD}{\underline{D}}
\newcommand{\uv}{\underline{v}}
\newcommand{\uT}{\underline{T}}
\newcommand{\ut}{\underline{t}}
\newcommand{\ur}{\underline{r}}
\newcommand{\uR}{\underline{R}}
\newcommand{\uc}{\underline{c}}
\newcommand{\uC}{\underline{C}}
\newcommand{\ul}{\underline{l}}
\newcommand{\uL}{\underline{L}}
\newcommand{\uh}{\underline{h}}
\newcommand{\uH}{\underline{H}}
\newcommand{\ue}{\underline{e}}
\newcommand{\uE}{\underline{E}}
\newcommand{\uG}{\underline{G}}
\newcommand{\ug}{\underline{g}}
\newcommand{\uz}{\underline{z}}
\newcommand{\uZ}{\underline{Z}}
\newcommand{\uu}{\underline{u}}
\newcommand{\uU}{\underline{U}}
\newcommand{\uj}{\underline{j}}
\newcommand{\uJ}{\underline{J}}
\newcommand{\uX}{\underline{X}}
\newcommand{\ux}{\underline{x}}
\newcommand{\uY}{\underline{Y}}
\newcommand{\uy}{\underline{y}}
\newcommand{\uW}{\underline{W}}
\newcommand{\uw}{\underline{w}}
\newcommand{\uth}{\underline{\theta}}
\newcommand{\uTh}{\underline{\Theta}}
\newcommand{\uph}{\underline{\phi}}
\newcommand{\ual}{\underline{\alpha}}
\newcommand{\uxi}{\underline{\xi}}
\newcommand{\us}{\underline{s}}
\newcommand{\uS}{\underline{S}}
\newcommand{\un}{\underline{n}}
\newcommand{\uN}{\underline{N}}
\newcommand{\up}{\underline{p}}
\newcommand{\uq}{\underline{q}}
\newcommand{\uf}{\underline{f}}
\newcommand{\ua}{\underline{a}}
\newcommand{\ub}{\underline{b}}
\newcommand{\uDelta}{\underline{\Delta}}

\newcommand{\cA}{{\cal A}}
\newcommand{\tcA}{\tilde{\cA}}
\newcommand{\cB}{{\cal B}}
\newcommand{\cC}{{\cal C}}
\newcommand{\cc}{{\cal c}}
\newcommand{\cD}{{\cal D}}
\newcommand{\cE}{{\cal E}}
\newcommand{\cF}{{\cal F}}
\newcommand{\cH}{{\cal H}}
\newcommand{\cI}{{\cal I}}
\newcommand{\cK}{{\cal K}}
\newcommand{\cL}{{\cal L}}
\newcommand{\cN}{{\cal N}}
\newcommand{\cP}{{\cal P}}
\newcommand{\cQ}{{\cal Q}}
\newcommand{\cR}{{\cal R}}
\newcommand{\cS}{{\cal S}}
\newcommand{\cs}{{\cal s}}
\newcommand{\cT}{{\cal T}}
\newcommand{\ct}{{\cal t}}
\newcommand{\cU}{{\cal U}}
\newcommand{\cV}{{\cal V}}
\newcommand{\cW}{{\cal W}}
\newcommand{\cX}{{\cal X}}
\newcommand{\cx}{{\cal x}}
\newcommand{\cY}{{\cal Y}}
\newcommand{\cy}{{\cal y}}
\newcommand{\cZ}{{\cal Z}}
\newcommand{\tA}{\tilde{A}}
\newcommand{\tE}{\tilde{E}}
\newcommand{\tZ}{\tilde{Z}}
\newcommand{\tz}{\tilde{z}}
\newcommand{\tQ}{\tilde{Q}}
\newcommand{\tR}{\tilde{R}}
\newcommand{\hX}{\hat{X}}
\newcommand{\hY}{\hat{Y}}
\newcommand{\hZ}{\hat{Z}}
\newcommand{\huX}{\hat{\uX}}
\newcommand{\huY}{\hat{\uY}}
\newcommand{\huZ}{\hat{\uZ}}
\newcommand{\inthalf}[1]{\int_{-\frac{1}{2}}^{\frac{1}{2}} #1 df}
\newcommand{\indp}{\underline{\; \| \;}}
\newcommand{\diag}{\mbox{diag}}
\newcommand{\sumk}{\sum_{k=1}^{K}}
\newcommand{\beq}[1]{\begin{equation}\label{#1}}
\newcommand{\eeq}{\end{equation}}
\newcommand{\req}[1]{(\ref{#1})}
\newcommand{\beqn}[1]{\begin{eqnarray}\label{#1}}
\newcommand{\eeqn}{\end{eqnarray}}
\newcommand{\limn}{\lim_{n \rightarrow \infty}}
\newcommand{\limN}{\lim_{N \rightarrow \infty}}
\newcommand{\limr}{\lim_{r \rightarrow \infty}}
\newcommand{\limd}{\lim_{\delta \rightarrow \infty}}
\newcommand{\limM}{\lim_{M \rightarrow \infty}}
\newcommand{\limsupn}{\limsup_{n \rightarrow \infty}}
\newcommand{\imii}{\int_{-\infty}^{\infty}}
\newcommand{\imix}{\int_{-\infty}^x}
\newcommand{\ioi}{\int_o^\infty}
\newcommand{\bphi}{\mbox{\boldmath \begin{math}\phi\end{math}}}
\newcommand{\bth}{\mbox{\boldmath \begin{math}\theta\end{math}}}
\newcommand{\bhth}{\mbox{\boldmath \begin{math}\hat{\theta}\end{math}}}
\newcommand{\bg}{\mbox{\boldmath \begin{math}g\end{math}}}
\newcommand{\bA}{{\bf A}}
\newcommand{\ba}{{\bf a}}
\newcommand{\bB}{{\bf B}}
\newcommand{\bb}{{\bf b}}
\newcommand{\bc}{{\bf c}}
\newcommand{\bd}{{\bf d}}
\newcommand{\bD}{{\bf D}}
\newcommand{\bff}{{\bf f}}
\newcommand{\bG}{{\bf G}}
\newcommand{\bW}{{\bf W}}
\newcommand{\bM}{{\bf M}}
\newcommand{\bi}{{\bf i}}
\newcommand{\bl}{{\bf l}}
\newcommand{\bm}{{\bf m}}
\newcommand{\bn}{{\bf n}}
\newcommand{\bp}{{\bf p}}
\newcommand{\bs}{{\bf s}}
\newcommand{\bt}{{\bf t}}
\newcommand{\bu}{{\bf u}}
\newcommand{\bx}{{\bf x}}
\newcommand{\by}{{\bf y}}
\newcommand{\bz}{{\bf z}}
\newcommand{\bC}{{\bf C}}
\newcommand{\bI}{{\bf I}}
\newcommand{\bJ}{{\bf J}}
\newcommand{\bN}{{\bf N}}
\newcommand{\bS}{{\bf S}}
\newcommand{\bT}{{\bf T}}
\newcommand{\bU}{{\bf U}}
\newcommand{\bV}{{\bf V}}
\newcommand{\bv}{{\bf v}}
\newcommand{\bw}{{\bf w}}
\newcommand{\bX}{{\bf X}}
\newcommand{\bY}{{\bf Y}}
\newcommand{\bZ}{{\bf Z}}
\newcommand{\bzero}{{\bf 0}}
\newcommand{\oI}{\overline{I}}
\newcommand{\oD}{\overline{D}}
\newcommand{\oh}{\overline{h}}
\newcommand{\oV}{\overline{V}}
\newcommand{\oR}{\overline{R}}
\newcommand{\oH}{\overline{H}}
\newcommand{\ol}{\overline{l}}
\newcommand{\E}{{\cal E}_d}
\newcommand{\thref}[1]{Theorem \ref{#1}}
\newcommand{\figref}[1]{Figure \ref{#1}}
\newcommand{\secref}[1]{Section \ref{#1}}

\newcommand{\round}{\mathop{\mathrm{round}}}
\newcommand{\var}{\mathop{\mathrm{Var}}}

\newcommand{\Nu}{{\mathcal V}}

\newcommand{\ej}{{e^{j2\pi f}}}
\newcommand{\WZ}{{\text{WZ}}}
\newcommand{\DPC}{{\text{DPC}}}
\newcommand{\SDR}{{\text{SDR}}}
\newcommand{\SNR}{{\text{SNR}}}
\newcommand{\tSNR}{\widetilde\SNR}

\newcommand{\yesindent}{\hspace*{\parindent}}   
\newcommand{\pderiv}[2]{\frac{ \partial {#1}}{ \partial {#2}}}
\newcommand{\overr}[2]{\left({\begin{array}{l}#1\\#2\end{array}}\right)}
\newcommand{\Ddef}{\stackrel{\Delta}{=}}

\pagestyle{plain}

\newcommand{\bQ}{{\bf Q}}
\newcommand{\bq}{{\bf q}}
\newcommand{\el}{\ell}
\newcommand{\linf}{{\el\rightarrow\infty}}
\renewcommand{\thesection}{\Roman{section}}
\renewcommand{\thesubsection}{\thesection-\Alph{subsection}}

\newcommand{\rem}[1]{}

\newtheorem{theorem}{Theorem}
\newtheorem{prop}{Proposition}
\newtheorem{cor}{Corollary}
\newtheorem{lemma}{Lemma}
\newtheorem{conj}{Conjucture}
\newtheorem{assume}{Assumption}

\newcommand{\ccaption}{\caption*~{Figure~\thefigure: }}


\title{Joint Wyner-Ziv/Dirty-Paper Coding by Modulo-Lattice
Modulation\symbolfootnote[2]{Parts of this work were presented at
ISIT2006, Seattle, WA, July 2006. This work was supported by the
Israeli Science Foundation (ISF) under grant \# 1259/07, and by the
Advanced Communication Center (ACC). The first author was also
supported by a fellowship of the Yitzhak and Chaya Weinstein
Research Institute for Signal Processing at Tel Aviv University. }}

\author{Yuval Kochman and Ram Zamir \\
Dept. Electrical Engineering - Systems, Tel Aviv University}

\maketitle

\begin{abstract}
The combination of source coding with decoder side-information
(Wyner-Ziv problem) and channel coding with encoder side-information
(Gel'fand-Pinsker problem) can be optimally solved using the
separation principle. In this work we show an alternative scheme for
the quadratic-Gaussian case, which merges source and channel coding.
This scheme achieves the optimal performance by a applying
modulo-lattice modulation to the analog source. Thus it saves the
complexity of quantization and channel decoding, and remains with
the task of ``shaping'' only. Furthermore, for high signal-to-noise
ratio (SNR), the scheme approaches the optimal performance using an
SNR-independent encoder, thus it is robust to unknown SNR at the
encoder.
\end{abstract}

\vspace{5mm} \textbf{keywords:} joint source/channel coding, analog
transmission, Wyner-Ziv problem, writing on dirty paper, modulo
lattice modulation, MMSE estimation, unknown SNR, broadcast channel.

\section{Introduction} \label{Sec_Intro}


\begin{figure}[t]
\centering
\input{joint_prob2.pstex_t}
\ccaption{The Wyner-Ziv / dirty-paper coding problem}
\label{joint_prob_fig}
\end{figure}

Consider the quadratic-Gaussian joint source/channel coding problem
for the Wyner-Ziv (WZ) source \cite{WynerZiv76} and Gel'fand-Pinsker
channel \cite{GelfandPinsker80}, as depicted in
\figref{joint_prob_fig}. In the Wyner-Ziv setup, the source is
jointly distributed with some side information (SI) known at the
decoder. In the Gaussian case, the WZ-source sequence $S_k$ is given
by: \beq{WZ_source} S_k = Q_k + J_k \ \ , \eeq where the unknown
source part, $Q_k$, is Gaussian i.i.d. with variance $\sigma_Q^2$,
while $J_k$ is an arbitrary SI sequence known at the decoder. In the
Gel'fand-Pinsker setup, the channel transition distribution depends
on a state that serves as encoder SI. In the Gaussian case, known as
the dirty paper channel (DPC) \cite{Costa83}, the DPC output, $Y_k$,
is given by: \beq{WDP_channel} Y_k = X_k + Z_k + I_k \ \ , \eeq
where $X_k$ is the channel input, the unknown channel noise, $Z_k$,
is Gaussian i.i.d. with variance $N$, while $I_k$ is an arbitrary
interference, known at the encoder. When referring to $I_k$ and
$J_k$, we use the terms interference and SI interchangeably, since
they may be seen either as external components added to the source
and to the channel noise, or as known parts of these entities. From
here onward we use the bold notation to denote $K$-dimensional
vectors, i.e. \[ \bX = [X_1,\cdots,X_k,\cdots,X_K] \ \ . \] The
sequences $\bQ,\bJ,\bZ$~and~$\bI$ are all mutually independent,
hence the channel noise $\bZ$ is independent of the channel input
sequence $\bX$. The encoder is some function of the source vector
that may depend on the channel SI vector as well:
\beq{generic_encoder} \bX = f(\bS,\bI) \ \ , \eeq and must obey the
power constraint \beq{power} \frac{1}{K} E\{ \|\bX\|^2 \} \leq P ,
\eeq where $\|\cdot\|$ denotes the Euclidean norm. The decoder is
some function of the channel output vector that may depend on the
source SI vector as well: \beq{generic_decoder} \hat \bS =
g(\bY,\bJ) \ \ , \eeq and the reconstruction quality performance
criterion is the mean-squared error (MSE): \beq{distortion} D =
\frac{1}{K} E\{ \| \hat \bS - \bS \| ^2 \} \ \ . \eeq

The setup of \figref{joint_prob_fig} described above is a special
case of the joint WZ-source and Gel'fand-Pinsker channel setting.
Thus, by Merhave and Shamai \cite{MerhavShamaiWZGP}, Shannon's
separation principle holds. So a combination of optimal source and
channel codes can approach the optimum distortion $D^{opt}$,
satisfying: \beq{OPTA} R_\WZ(D^{opt}) = C_\DPC \eeq where $R_\WZ(D)$
is the WZ-source rate-distortion function and $C_\DPC$ is the dirty
paper channel capacity. However, the optimality of ``digital''
separation-based schemes comes at the price of large delay and
complexity. Moreover, they suffer from lack of robustness: if the
channel signal-to-noise ratio (SNR) turns out to be lower than
expected, the resulting distortion may be very large, while if the
SNR is higher than expected, there is no improvement in the
distortion \cite{Ziv70,TrottITW96}.

In the special case of white Gaussian source and channel without
side information ($\bI=\bJ=\bzero$), it is well known that analog
transmission is optimal \cite{Goblick65}. In that case, the encoding
and decoding functions \beq{Goblick} \begin{array}{ccc}
                                       X_k & = & \beta S_k \ \ , \\
                                       \hat S_k & = & \frac{\alpha}{\beta} Y_k
                                     \end{array} \eeq
are mere scalar factors, where $\beta$ is a ``zoom in'' factor
chosen to satisfy the channel power constraint and $\alpha$ is the
channel MMSE (Wiener) coefficient. This scheme achieves the optimal
distortion \eqref{OPTA} while having low complexity (two
multiplications per sample), zero delay and \emph{full robustness}:
only the receiver needs to know the channel SNR, while the
transmitter is completely ignorant of that. Such a perfect matching
of the source to the channel, which allows \emph{single-letter
coding}, only occurs under very special conditions \cite{ToCode}.

In the quadratic-Gaussian setting in the presence of side
information, these conditions do not hold \cite{MerhavShamaiWZGP}.
It is interesting to note that in this case, $R_\WZ(D)$ is just the
Gaussian rate-distortion function for the unknown source part $\bQ$
\cite{Wyner78}, while $C_\DPC$ is just the AWGN capacity for the
channel noise $\bZ$ \cite{Costa83}, i.e. the SI components $\bI$ and
$\bJ$ are ``eliminated'' as would be done had they been known to
both the encoder and the decoder. We see, then, that this perfect
interference cancelation is not achievable by single-letter coding.

In this work we propose a scheme for the joint Wyner-Ziv/dirty-paper
problem that takes a middle path, i.e., a ``semi-analog'' solution
which partially gains the complexity and robustness advantages of
analog transmission: It can be made optimal (in the sense of
\eqref{OPTA}) for any fixed SNR, with reduced complexity. Moreover,
it allows a good compromise between the performance at different
SNRs, and becomes SNR-independent at the limit of high SNR.

The scheme we present subtracts the channel interference $\bI$ at
the encoder modulo-lattice, then uses again subtraction of the
source known part $\bJ$ in conjunction with modulo-lattice
arithmetic at the decoder. Thus it achieves an \emph{equivalent
single-letter channel} with $\bI=\bJ=\bzero$. Since the processing
is applied to the analog signal, without using any
information-bearing code, we call this approach \emph{modulo-lattice
modulation} (MLM).

Modulo-lattice codes were suggested as a tool for side information
source and channel problems; see
\cite{RamiShamaiUriLattices,BarronChenWornell}, where a lattice is
used for shaping of a digital code (which may itself have a lattice
structure as well, yielding a nested lattice structure).
Modulo-lattice transmission of an analog signal in the WZ setting
was first introduced in \cite{TzvikaBroadcast}, in the context of
joint source/channel coding with bandwidth expansion, i.e. when
there are several channel uses per each source sample. Here we
generalize and formalize this approach, and apply it to SI problems.
In a preliminary version of this work \cite{AnalogMatchingISIT}, we
used the MLM scheme as a building block in \emph{Analog Matching} of
colored sources to colored channels. Later, Wilson et al.
\cite{WilsonNarayananCaire07,WilsonNarayananCaireIT} used
transmission of an analog signal modulo a \emph{random} code to
arrive at similar results. Recently, MLM was used in network
settings for computation over the Gaussian MAC \cite{NazerGastpar07}
or for coding for the colored Gaussian relay network
\cite{RematchForwardISIT}.

The rest of the paper is organized as follows: In
\secref{Sec_Lattice} we bring preliminaries about multi-dimensional
lattices, and discuss the existence of lattices that are
asymptotically suitable for joint WZ/DPC coding. In
\secref{Sec_Joint} we present the joint WZ/DPC scheme and prove its
optimality. In \secref{Sec_Universal} we examine the scheme in an
unknown SNR setting and show its asymptotic robustness. Finally,
\secref{conclusion} discusses complexity reduction issues.

\section{Background: Good Shaping Lattices for Analog Transmission}
\label{Sec_Lattice}

Before we present the scheme, we need some definitions and results
concerning multi-dimensional lattices. Let $\Lambda$ be a
$K$-dimensional lattice, defined by the generator matrix $G \in
\mathbb{R}^{K \times K}$. The lattice includes all points
$\{\bl=G\cdot\bi:\bi\in\mathbb{Z}^K\}$ where $\mathbb{Z}=\{0,\pm 1,
\pm2, \ldots\}$. The nearest neighbor quantizer associated with
$\Lambda$ is defined by \[Q(\bx) = \arg\min_{\bl\in\Lambda}
\|\bx-\bl\| \ \ ,\] where $\|\cdot\|$ denotes the Euclidian norm and
ties are broken in a systematic manner. Let the basic Voronoi cell
of $\Lambda$ be
\[\Nu_0 = \{\bx : Q(\bx)=\textbf{0}\}\ \ .\] The second
moment of a lattice is given by the variance of a uniform
distribution over the basic Voronoi cell: \beq{lattice_power}
\sigma^2(\Lambda) = \frac{1}{K} \int_{\Nu_0} \|\bx\|^2 d\bx \ \ .
\eeq The modulo-lattice operation is defined by:
\[\bx \bmod{\Lambda} = \bx - Q(\bx)\ \ .\] By definition, this
operation satisfies the ``distributive law'': \beq{mod_mod} [\bx
\bmod{\Lambda} + \by] \bmod{\Lambda} = [\bx+\by]\bmod{\Lambda} \ \
.\eeq The covering radius of a lattice is given by
\beq{covering_radius} r(\Lambda) = \max_{\bx \in \Nu_0} \|\bx\| \ \
. \eeq

For a dither vector $\bd$, the dithered modulo-lattice operation is:
\[\by = [\bx+\bd] \bmod{\Lambda} \ \ . \] If the dither vector $\bD$
is independent of $\bx$ and uniformly distributed over the basic
Voronoi cell $\Nu_0$, then $\bY=[\bx+\bD]\bmod\Lambda$ is uniformly
distributed over $\Nu_0$ as well, and independent of $\bx$
\cite{FederZamirLQN}. Consequently, the second moment of $\bY$ per
element is $\sigma^2(\Lambda)$.

The loss factor $L(\Lambda,p_e)$ of a lattice w.r.t. Gaussian noise
at error probability $p_e$ is defined as follows. Let $\bZ$ be
Gaussian i.i.d. vector with element variance equal to the lattice
second moment $\sigma^2(\Lambda)$. Then \beq{simple_L_def}
L(\Lambda,p_e) = \min\left\{l : \ \Pr\left\{\frac{\bZ}{\sqrt{l}}
\notin \Nu_0 \right\} \leq p_e \right\} \ \ . \eeq For small enough
$p_e$ this factor is at least one. By \cite[Theorem
5]{GoodLattices}, there exists a sequence of lattices which
possesses a vanishing loss at the limit of high
dimension\footnote{These lattices are simultaneously good for source
and channel coding; see more on this in
Appendix~\ref{appendix_lattice}.}, i.e.: \beq{good_L}
\lim_{p_e\rightarrow 0} \lim_{K\rightarrow\infty} L(\Lambda_K,p_e) =
1 \ \ . \eeq Moreover, there exists a sequence of such lattices that
is also \emph{good for covering}, i.e. defining: \beq{tilde_L_def}
\tilde L (\Lambda) = \frac{r^2(\Lambda)}{K \cdot \sigma^2(\Lambda)}
\ \ , \eeq where $r(\Lambda)$ was defined in
\eqref{covering_radius}, the sequence also satisfies\footnote{Note
that by definition, $\tilde L(\Lambda_K)\geq 1$ always.}:
$\lim_{K\rightarrow\infty} \tilde L(\Lambda_K) = 1$. However, for
this work we need a slightly modified result, which allows to
replace the Gaussian noise by a combination of Gaussian and
``self-noise'' components. To that end, we define for any
$0\leq\alpha\leq 1$ the $\alpha$-mixture noise as:
\[\bZ_{\alpha} = \sqrt{1-(1-\alpha)^2} \bW - (1-\alpha) \bD\ \ ,\]
where $\bW$ is Gaussian i.i.d. with element variance
$\sigma^2(\Lambda)$, and $\bD$ is uniform over $\Nu_0$ and
independent of $\bW$. Note that since
$\frac{1}{K}\|\bD\|^2=\sigma^2(\Lambda)$, the resulting mixture also
has average per-element variance $\sigma^2(\Lambda)$. We re-define
the loss factor w.r.t. this mixture noise as \beq{L_def}
L(\Lambda,p_e,\alpha) = \min\left\{l : \
\Pr\left\{\frac{\bZ_\alpha}{\sqrt{l}} \notin \Nu_0 \right\} \leq p_e
\right\} \ \ . \eeq Note that this definition reduces to
\eqref{simple_L_def} for $\alpha=1$. Using this definition, we have
the following, which is a direct consequence of \cite{UriRamiAWGN}.

\vspace{5mm}
\begin{prop} \label{prop_lattice} \textbf{(Existence of good lattices)}
For any error probability $p_e>0$, and for any $0\leq\alpha\leq 1$,
there exists a sequence of $K$-dimensional lattices $\Lambda_K$
satisfying: \beq{prop_1} \lim_{p_e\rightarrow 0}
\lim_{K\rightarrow\infty} L(\Lambda_K,p_e,\alpha) = 1 \ \ , \eeq and
\beq{prop_2} \lim_{K\rightarrow\infty} \tilde L(\Lambda_K) = 1 \ \ .
\eeq
 \vspace{5mm}
\end{prop}

Note that since by definition, $L(\Lambda_K,p_e,\alpha)$ is
non-increasing in $p_e$, it follows that for \emph{any} $p_e>0$ this
sequence of lattices satisfies: \beq{limsup}
\limsup_{K\rightarrow\infty} L(\Lambda_K,p_e,\alpha) \leq 1 \ \ .
\eeq In Appendix~\ref{appendix_lattice} we elaborate more on the
significance of this result, and on its connection to more commonly
used measures of goodness of lattices.

\section{Modulo-Lattice WZ/DPC Coding} \label{Sec_Joint}

We now present the joint source/channel scheme for the SI problem of
\figref{joint_prob_fig}. As explained in the Introduction, the
quadratic-Gaussian rate-distortion function (RDF) of the WZ source
\eqref{WZ_source} is equal to the RDF of the source $Q_k$ (without
the known part $J_k$), given by: \beq{RDF} R_\WZ(D) =
\frac{1}{2}\log\frac{\sigma_Q^2}{D} \ \ . \eeq Similarly, the
capacity of the Gaussian DPC \eqref{WDP_channel} is equal to the
AWGN capacity (without the interference $I_k$): \beq{C} C_\DPC =
\frac{1}{2}\log\left(1+\frac{P}{N}\right) \ \ . \eeq Recalling that
the separation principle holds for this problem
\cite{MerhavShamaiWZGP}, the optimum distortion \eqref{OPTA} is thus
given by: \beq{joint_optimal} D^{opt} = \frac{N}{P+N} \sigma^2_Q \ \
\ . \eeq

\begin{figure*}[!t]
\centering
\input{joint_scheme4.pstex_t}
\ccaption{Analog Wyner-Ziv / dirty-paper coding scheme: $\bS$ =
source, $\hat\bS$ = reconstruction, $\bZ$ = channel noise, $\bI$ =
interference known at the encoder, $\bJ$ = source component known at
the decoder, $\bD$ = dither} \label{joint_scheme_fig}
\end{figure*}

We show how to approach $D^{opt}$ using the joint source/channel
coding scheme depicted in \figref{joint_scheme_fig}. In this scheme,
the $K$-dimensional encoding and decoding functions
\eqref{generic_encoder},\eqref{generic_decoder} are given by:
\begin{subequations} \label{joint_encdec}
\begin{align} \label{joint_encoder} \bX =& [\beta \bS + \bD -
\alpha\bI] \bmod{\Lambda} \\
\label{joint_decoder} \hat \bS =&
\frac{\alpha_S}{\beta}\Bigl\{[\alpha_C \bY - \bD - \beta\bJ]
\bmod{\Lambda} \Bigr\} + \bJ \ \ ,
\end{align} \end{subequations}
respectively, where the second moment \eqref{lattice_power} of the
lattice is $\sigma^2(\Lambda)=P$, and the dither vector $\bD$ is
uniformly distributed over $\Nu_0$ and independent of the source and
of the channel. The channel power constraint is satisfied
automatically by the properties of dithered lattice quantization
discussed in \secref{Sec_Lattice}. The factors $\alpha_S$,
$\alpha_C$ and $\beta$ will be chosen in the sequel. For optimum
performance, $\beta$ which is used at the encoder will depend upon
the variance of the source unknown part, while $\alpha_C$ used at
the decoder will depend upon the channel SNR. It is assumed, then,
that both the encoder and the decoder have full knowledge of the
source and channel statistics; we will break with this assumption in
the next section.

The following theorem gives the performance of the scheme, in terms
of the lattice parameters $L(\cdot,\cdot,\cdot)$ in \eqref{L_def}
and in $\tilde L(\cdot)$  \eqref{tilde_L_def}, and the quantities:
\begin{subequations} \label{params_def}
\begin{align}
\label{alpha_0} \alpha_0 \Ddef & \frac{P}{P+N}  ,\\
\label{tilde_alpha} \tilde \alpha \Ddef& \max\left(\alpha_0 -
\frac{L(\Lambda,p_e,\alpha_0)-1}{L(\Lambda,p_e,\alpha_0)},0\right) \
\ .
\end{align}
\end{subequations} We will also use these quantities in the sequel
to specify the choice of factors $\alpha_S$, $\alpha_C$ and $\beta$.

\vspace{5mm}
\begin{theorem} \label{thm_any_lattice} \textbf{(Performance of the MLM scheme with any lattice)}
For any lattice $\Lambda$ and any error probability $p_e>0$, there
exists a choice of factors $\alpha_C,\alpha_S,\beta$ such that the
system of \eqref{joint_encdec} (depicted in
\figref{joint_scheme_fig}) satisfies:
\[ D \leq L(\Lambda,p_e,\alpha_0) D^{opt} +
p_e D^{max} \ \ , \] where the optimum distortion $D^{opt}$ was
defined in \eqref{joint_optimal}, and \beq{d_max} D^{max} = 4
\sigma_Q^2 \left(1 + \frac{\tilde L(\Lambda)}{\tilde \alpha} \right)
\ \ . \eeq
\end{theorem}
\vspace{5mm}

We prove this theorem in the sequel. As a direct corollary from it,
taking $p_e$ to be an arbitrarily small probability and using the
properties of good lattices \eqref{prop_2} and \eqref{limsup}, we
have the following asymptotic optimality result\footnote{The
explicit derivation of $D^{max}$ is not necessary for proving
\thref{thm_joint}; see Appendix~\ref{Appendix_Wyner}.}

\vspace{5mm}
\begin{theorem} \label{thm_joint} \textbf{(Optimality of the MLM scheme)}
Let $D(\Lambda_K)$ be the distortion achievable by the system of
\eqref{joint_encdec} with a lattice from a sequence $\{\Lambda_K\}$
that is simultaneously good for source and channel coding in the
sense of Proposition~\ref{prop_lattice}. Then for any $\epsilon>0$,
there exists a choice of factors $\alpha_C$, $\alpha_S$ and $\beta$,
such that
\[\limsup_{K\rightarrow\infty} D\left(\Lambda_K\right) \leq D^{opt} + \epsilon \ \ . \]
\vspace{3mm}
\end{theorem}

For proving \thref{thm_any_lattice} we start with a lemma, showing
equivalence in probability to a real-additive noise channel (see
\figref{output_eq_fig}). The equivalent additive noise is:
\beq{Z_eq} \bZ_{eq} = \alpha_C \bZ - (1-\alpha_C) \bX \ \ , \eeq
where $\bZ$ and $\bX$ are the physical channel input and AWGN,
respectively. By the properties of the dithered modulo-lattice
operation, the physical channel input $\bX$ is uniformly distributed
over $\Nu_0$ and independent of the source. Thus, $\bZ_{eq}$ is
indeed additive and has per-element variance: \beq{sigma_eq}
\sigma_{eq}^2 = \alpha_C^2 N + (1-\alpha_C)^2 P \ \ . \eeq

\begin{figure}[!t]
\centering \subfloat[Equivalent modulo-lattice channel.]
    {\label{finite_k_eq_fig}
      \input{finite_k_equivalent2.pstex_t}}
      \\ \vspace{2mm}
      \subfloat[Equivalent real-additive noise channel w.p. $(1-p_e)$.]
      {\label{output_eq_fig}
      \input{output_equivalent4.pstex_t}}
      \ccaption{Equivalent channels for the WZ/WDP coding scheme}
\label{joint_equivalent_fig}
\end{figure}

\vspace{5mm}
\begin{lemma} \label{lemma_joint} \textbf{(Equivalent additive noise
channel)} Fix some $p_e>0$. In the system defined by
\eqref{WZ_source},\eqref{WDP_channel} and \eqref{joint_encdec}, the
decoder modulo output $\bM$ (see \figref{joint_scheme_fig})
satisfies: \beq{M} \bM = \beta\bQ + \bZ_{eq} \ \ \mbox{w.p.
$(1-p_e)$,} \eeq provided that \beq{joint_condition} \beta^2
\sigma_Q^2 + \sigma_{eq}^2 \leq \frac{P}{L(\Lambda,p_e,\alpha_C)} \
\ , \eeq where $\bZ_{eq}$, defined in \eqref{Z_eq}, is independent
of $\bQ$ and $\bJ$ and has per-element variance $\sigma_{eq}^2$
\eqref{sigma_eq}, and $L(\cdot,\cdot,\cdot)$ was defined in
\eqref{L_def}. \vspace{3mm} \end{lemma}

Consequently, as long as \eqref{joint_condition} holds, the whole
system is equivalent with probability $(1-p_e)$ to the channel
depicted in \figref{output_eq_fig}: \beqn{joint_eq_2} \hat \bS &=&
\bJ + \frac{\alpha_S}{\beta} \bZ_{eq} + \alpha_S \bQ \nonumber
\\ &=& \bS + \frac{\alpha_S}{\beta} \bZ_{eq} - (1-\alpha_S)
\bQ \ \ . \eeqn

\begin{proof}
We will first prove equivalence to the channel of
\figref{finite_k_eq_fig}: \beq{joint_eq_1} \bM = [\beta \bQ
+\bZ_{eq}] \bmod{\Lambda} \ \ , \eeq where $\bZ_{eq}$ was defined in
\eqref{Z_eq}. To that end, let $\bT=\alpha_C\bY-\bD-\beta\bJ$ denote
the input of the decoder modulo operation (see \eqref{joint_decoder}
and \figref{joint_scheme_fig}). Combine \eqref{WDP_channel} and
\eqref{joint_encoder} to assert: \beqn{T_joint_proof} \bT &=&
\alpha_C
(\bX+\bZ+\bI)- \bD - \beta\bJ \nonumber \\
&=& [\beta \bS + \bD - \alpha_C \bI] \bmod{\Lambda} + \bZ_{eq}
+\alpha_C \bI- \bD - \beta\bJ \ \ . \nonumber \eeqn Now, using
\eqref{WZ_source} and the ``distributive law'' \eqref{mod_mod}:
\[\bT\bmod{\Lambda} = [\beta \bQ +\bZ_{eq}] \bmod{\Lambda} \ \ , \] and
since $\bT = \bM \bmod \Lambda$, we establish \eqref{joint_eq_1}.
Now we note that  \[ \beta\bQ+\bZ_{eq} = \beta\bQ + \alpha_C \bZ -
(1-\alpha_C) \bX \Ddef \sqrt{1-(1-\alpha_C)^2} \bW - (1-\alpha_C)
\bX \ \ ,
\] where $\bW$ is Gaussian i.i.d., $\bX$ is uniform over the basic
cell $\Nu_0$ of the lattice $\Lambda$, and the total variance (per
element) is given by the l.h.s. of \eqref{joint_condition}. By the
definition of $L(\cdot,\cdot,\cdot)$, we have that \beq{in_cell} \bT
= \beta\bQ+\bZ_{eq} \in \Nu_0 \eeq w.p. at least $(1-p_e)$.
Substituting this in \eqref{joint_eq_1}, we get \eqref{M}.
\end{proof}

This channel equivalence holds for any choice of dimension $K$,
lattice $\Lambda$ and factors $\alpha_C$, $\alpha_S$ and $\beta$, as
long as \eqref{joint_condition} holds. For the proof of
\thref{thm_any_lattice} we make the following choice (using the
parameters of \eqref{params_def}):
\begin{subequations} \label{finite_K_parameters}
\begin{align}
\label{alpha_C} \alpha_C =& \alpha_0 \ \ ,\\
\label{beta} \beta^2 =& \tilde \alpha \frac{P}{\sigma_Q^2} \ \ ,
\\
\label{alpha_S} \alpha_S =& \frac{\tilde\alpha P }{\tilde \alpha P +
\alpha_0 N} \ \ .
\end{align}
\end{subequations}
It will become evident in the sequel, that $\alpha_C$ and $\alpha_S$
are the MMSE (Wiener) coefficients for estimating $\bX$ from
$\bX+\bZ$ and $\bQ$ from $\bQ+\frac{\bZ_{eq}}{\beta}$, respectively,
while $\beta$ is the maximum zooming factor that allows to satisfy
\eqref{joint_condition} with equality, whenever possible.

\vspace{3mm} \emph{Proof of \thref{thm_any_lattice}}: For
calculating the achievable distortion, first note that by the
properties of MMSE estimation,
\[\sigma_{eq}^2 = \alpha_C N = \alpha_0 N \ \ . \] Using this, it can be verified
that our choice of $\beta$ satisfies \eqref{joint_condition}, thus
\eqref{joint_eq_2} holds with probability $(1-p_e)$. Denoting by
$D^{correct}$ and $D^{incorrect}$ the distortions when
\eqref{joint_eq_2} holds or does not hold, respectively, we have:
\beqn{D12} D &=& (1-p_e) D^{correct} + p_e D^{incorrect} \nonumber \\
&\leq& D^{correct} + p_e D^{incorrect} \ \ . \eeqn We shall now
bound both conditional distortions. For the first one, we have:
\beqn{D1} D^{correct} &=& \frac{1}{K}
E\left\{\left\|\frac{\alpha_S}{\beta_K}\bZ_{eq} -
(1-\alpha_S)\bQ\right\|^2\right\} \nonumber \\ &\stackrel{(a)}{=}&
\alpha_S \frac{\sigma_{eq}^2}{\beta^2} = \frac{\sigma_Q^2
\sigma_{eq}^2}{\beta^2\sigma_Q^2+\sigma_{eq}^2} \nonumber \\ &=&
\frac{D^{opt}}{1-\alpha_0+\tilde \alpha} \nonumber \\ &=&
\min\left(L(\Lambda,p_e,\alpha_C)D^{opt},\sigma_Q^2\right) \nonumber
\leq L(\Lambda,p_e,\alpha_C)D^{opt} \ \ , \eeqn where (a) stems from
the properties of MMSE estimation. It remains to show that
$D^{incorrect}\leq D^{max}$, which is established in
Appendix~\ref{appendix_D2}. $\Box$

As mentioned in the Introduction, a recent work
\cite{WilsonNarayananCaireIT} derives a similar asymptotic result,
replacing the shaping lattice of our scheme by a \emph{random}
shaping code. Such a choice is less restrictive since it is not tied
to the properties of good Euclidean lattices, though it leads to
higher complexity due to the lack of structure. The use of lattices
also allows analysis in finite dimension as in
\thref{thm_any_lattice} and in \secref{conclusion}. Furthermore,
structure is essential in network joint source/channel settings; see
e.g. \cite{NazerGastpar07}. Lastly, the dithered lattice formulation
allows to treat any interference signals, see Remark 2 in the
sequel.

We conclude this section by the following remarks, intended to shed
more light on the significance of the results above.

1. \textbf{Optimal decoding.} The decoder we described is \emph{not}
the MMSE estimator of $\bS$ from $\bY$. This is for two reasons:
First, the decoder ignores the probability of incorrect lattice
decoding. Second, since $\bZ_{eq}$ is not Gaussian, the
modulo-lattice operation w.r.t. the lattice Voronoi cells is not
equivalent to maximum-likelihood estimation of the lattice point
(see \cite{UriRamiAWGN} for a similar discussion in the context of
channel coding). Consequently, for any finite dimension the decoder
can be improved. We shall discuss further the issue of working with
finite-dimension lattices in \secref{conclusion}.

2. \textbf{Universality w.r.t. $\bI$ and $\bJ$.} None of the scheme
parameters depend upon the nature of the channel interference $\bI$
and source known part $\bJ$. Consequently, the scheme is adequate
for arbitrary (individual) sequences. This has no effect on the
asymptotic performance of \thref{thm_joint}, but for
finite-dimensional lattices the scheme may be improved, e.g. if the
interference signals are known to be Gaussian with low enough
variance. A similar argument also holds when the source or channel
statistics is not perfectly known, see \secref{Sec_Universal} in the
sequel.

3. \textbf{Non-Gaussian Setting.} If the source unknown part $\bQ$
or the channel noise $\bZ$ are not Gaussian, the optimum
quadratic-Gaussian distortion $D^{opt}$ may still be approached
using the MLM scheme, though it is no longer the optimum performance
for the given source and channel.

4. \textbf{Asymptotic choice of parameters.} In the limiting case
where $L(\Lambda,p_e,\alpha_0)\rightarrow 1$, we have that
$\alpha_S=\tilde \alpha = \alpha_0$ in \eqref{finite_K_parameters},
i.e. the choice of parameters approaches:
\begin{subequations} \label{Limit_parameters}
\begin{align}
\label{Limit_alpha} \alpha_C = \alpha_S =& \frac{P}{P+N} = \alpha_0 \ \ ,\\
\label{Limit_beta} \beta^2 =& \alpha_0 \frac{P}{\sigma_Q^2} \ \ .
\end{align}
\end{subequations}

5. \textbf{Properties of the equivalent additive-noise channel.}
With high probability, we have the equivalent real-additive noise
channel of \eqref{joint_eq_2} and \figref{output_eq_fig}. This
differs from the modulo-additivity of the lattice strategies of
\cite{UriRamiAWGN,LatticeStrategies}: Closeness of point under a
modulo arithmetic does not mean closeness under a difference
distortion measure. The condition \eqref{joint_condition} forms an
output-power constraint: No matter what the noise level of the
channel is, its output must have a power of no more than $P$; this
replaces the input-power constraint of the physical channel.
Furthermore, by the lattice quantization noise properties
\cite{FederZamirLQN}, the ``self noise'' component $(1-\alpha_C)\bX$
in \eqref{Z_eq} is asymptotically Gaussian i.i.d., and consequently
so is the equivalent noise $\bZ_{eq}$. Thus the additive equivalent
channel \eqref{joint_eq_2} is asymptotically an \emph{output-power
constrained AWGN channel}.

6. \textbf{Noise margin.} The additivity in \eqref{joint_eq_2} is
achieved through leaving a ``noise margin''.  The condition
\eqref{joint_condition} means that the sum of the (scaled) unknown
source part and equivalent noise should ``fit into'' the lattice
cell (see \eqref{in_cell}). Consequently, the unknown source part
$\bQ$ is inflated to a power strictly smaller than the lattice power
$P$. In the limit of infinite dimension, when the choice of
parameters becomes \eqref{Limit_parameters}, this power becomes
$\beta^2 \sigma_Q^2 = \alpha_0 P$. In comparison, it is shown in
\cite{LatticeStrategies} that in a lattice solution to a digital SI
problem, if the information-bearing code (fine lattice) occupies a
portion of power $\gamma P$ with any $\alpha_0 \leq \gamma \leq 1$,
capacity is achieved\footnote{In \cite{AmirSuperposition} a similar
observation is made, and a code of power $\alpha_0 P$ is presented
as a preferred choice, since it allows easy iterative decoding
between the information-bearing code and the coarse lattice.}. This
freedom, however, has to do with the modulo-additivity of the
equivalent channel; in our joint source/channel setting, necessarily
$\gamma=\alpha_0$.

7. \textbf{Comparison with analog transmission.} Lastly, consider
the similarity between our asymptotic AWGN channel and the optimal
analog transmission scheme without SI \eqref{Goblick}: Since we have
``eliminated from the picture'' the SI components $\bI$ and $\bJ$,
we are left with the transmission of the source unknown component
through an equivalent additive noise channel. As mentioned above,
the unknown source part $\bQ$ is only adjusted to power $\alpha_0 P$
(in the limit of high dimension), while in \eqref{Goblick} the
source $\bS$ is adjusted to power $P$; but since the equivalent
noise $\bZ_{eq}$ has variance $\alpha_0 N$, the equivalent channel
has signal-to-noise ratio of $P/N$, just as the physical channel.

\section{Transmission under Uncertainty Conditions}
\label{Sec_Universal}

We now turn to case where either the variance of the channel noise
$N$, or the variance of the source unknown part $\sigma_Q^2$, are
unknown at the encoder\footnote{We do not treat uncertainty at the
decoder, since $N$ can be learnt, while the major insight into the
matter of unknown $\sigma_Q^2$ is gained already by assuming
uncertainty at the encoder.}. In Section~\ref{sub_asymptotic} we
assume that $\sigma_Q^2$ is known at both sides, but the channel SNR
is unknown at the encoder. We show that in the limit of high SNR,
optimality can still be approached. In Section~\ref{sub_broadcast},
we address the general SNR case, as well as the case of unknown
$\sigma_Q^2$; for that, we adopt an alternative broadcast-channel
point of view.

For convenience, we present our results in terms of the channel
signal-to-noise ratio \beq{SNR} \SNR\Ddef\frac{P}{N} \eeq and the
achieved signal-to-distortion ratio \beq{SDR}
\SDR\Ddef\frac{\sigma_Q^2}{D} \ \ . \eeq Denoting the theoretically
optimal SDR as $\SDR^{opt}$, \eqref{joint_optimal} becomes:
\beq{SDR_opt} \SDR^{opt} = 1 + \SNR \ \ . \eeq

Our achievability results in this section are based upon application
of the MLM scheme, generally with a sub-optimal choice of parameters
due to the uncertainty. We only bring asymptotic results, using
high-dimensional ``good'' lattices. We present, then, the following
lemma, using the definition: \beq{beta_0} \beta_0^2 =
\frac{P}{\sigma_Q^2} \ \ . \eeq

\vspace{3mm}
\begin{lemma} \label{lemma_mismatch}
Let $\SDR(\Lambda_K)$ be the distortion achievable by the system of
\eqref{joint_encdec} with a lattice from a sequence $\{\Lambda_K\}$
that is good in the sense of Proposition~\ref{prop_lattice}. For any
choice of factors $\alpha_C$, $\alpha_S$ and $\beta$,
\beq{mismatch_SDR} \liminf_{K\rightarrow\infty} \SDR
\left(\Lambda_K\right) \geq \frac{\beta^2}{(1-\alpha_S)^2\beta^2 +
\alpha_S^2
\left[\frac{\alpha_C^2}{\SNR}+(1-\alpha_C)^2\right]\beta_0^2}\ \ ,
\eeq provided that \beq{mismatch_condition}
\frac{\beta^2}{\beta_0^2} + \frac{\alpha_C^2}{\SNR}+(1-\alpha_C)^2 <
1 \ \ . \eeq\vspace{3mm}
\end{lemma}

\begin{proof} This is a direct application of
Lemma~\ref{lemma_joint} and of \eqref{limsup}. First we fix some
$p_e>0$, and note that \eqref{mismatch_condition} is equivalent to
\eqref{joint_condition}. The SDR of the equivalent channel
\eqref{joint_eq_2}, at the limit
$L(\Lambda_K,p_e,\alpha_C)\rightarrow 1$ is then given by
\eqref{mismatch_SDR}. Then for $p_e\rightarrow 0$ the effect of
decoding errors vanishes, as shown in Appendix~\ref{Appendix_Wyner}
\end{proof}

Note, that by substituting the asymptotically optimal choice of
parameters \eqref{Limit_parameters} in \eqref{mismatch_SDR}, the
limit becomes $\SDR^{opt}$.

\subsection{Asymptotic Robustness for Unknown SNR}
\label{sub_asymptotic}

Imagine that we know that $\SNR\geq\SNR_0$, for some specific
$\SNR_0$, and that $\sigma_Q^2$ is known. Suppose that we set the
scheme parameters such that the correct decoding condition
\eqref{mismatch_condition} holds for $\SNR_0$. Since the variance of
the equivalent noise can only decrease with the $\SNR$, correct
lattice decoding will hold for any $\SNR\geq\SNR_0$, and we are left
with the equivalent additive-noise channel where the resulting SDR
is a strictly decreasing function of the $\SNR$. We use this
observation to derive an asymptotic result, showing that for high
SNR a \emph{single} encoder can approach optimality simultaneously
for all actual SNR. To that end, we replace the choice given in
\eqref{finite_K_parameters}, which leads to optimality at one SNR,
by the high-SNR choice $\alpha_C=\alpha_S=1$, where $\beta$ is
chosen to ensure correct decoding even at the minimal $\SNR_0$.

\vspace{5mm} \begin{theorem} \label{thm_robust_WZ}
\textbf{(Robustness at high SNR)} Let the source and channel be
given by \eqref{WZ_source} and \eqref{WDP_channel}, respectively.
Then for any $\epsilon>0$, there exists an
\underline{SNR-independent} sequence of encoding-decoding schemes
(each one achieving $\SDR_K$) that satisfies:
\begin{equation}
\label{WZ_almostOPTA} \liminf_{K\rightarrow\infty} \SDR_K \geq
(1-\epsilon)\SDR^{opt} \ \ ,
\end{equation}
for all sufficiently large (but finite) SNR. I.e.,
\eqref{WZ_almostOPTA} holds for all $\SNR\geq\SNR_0(\epsilon)$,
where $\SNR_0(\epsilon)$ is finite for all $\epsilon>0$.
\vspace{3mm}
\end{theorem}

A limit of a \emph{sequence} of schemes is needed in the theorem,
rather than a single scheme, since for any single scheme we have
$p_e>0$, thus the effect of incorrect decoding cannot be neglected
in the limit $\SNR\rightarrow\infty$ (meaning that the convergence
in Lemma~\ref{lemma_mismatch} in not uniform). If we restricted our
attention to SNRs bounded by some arbitrarily high value, a single
scheme would be sufficient.

\begin{proof} We use a sequence of MLM schemes with good lattices in the
sense of Proposition~\ref{prop_lattice}. If $\alpha_C=1$, then any
\[\beta^2 < \frac{\SNR_0-1}{\SNR_0}\cdot\beta_0^2\] satisfies the
condition \eqref{mismatch_condition} for $\SNR_0$, thus for any
$\SNR\geq\SNR_0$. Here we assume that $\SNR_0>1$, w.l.o.g. since we
can always choose $\SNR_0(\epsilon)$ of the theorem accordingly.
With this choice and with $\alpha_S=1$, we have by
Lemma~\ref{lemma_mismatch} that the SDR may approach (for any
$\SNR\geq\SNR_0$): \beq{in_proof_17} \frac{\beta^2}{\beta_0^2} \SNR
= \frac{\SNR_0-1}{\SNR_0}\cdot\SNR =
\frac{\SNR_0-1}{\SNR_0}\cdot\frac{\SNR}{\SNR+1} \cdot\SDR^{opt} \geq
\frac{\SNR_0-1}{\SNR_0+1} \cdot \SDR^{opt} \ \ . \nonumber \eeq Now
take $\epsilon = \frac{\SNR_0-1}{\SNR_0+1} -1$. Since
$\lim_{\SNR_0\rightarrow\infty} \epsilon = 0$, one may find $\SNR_0$
for any $\epsilon>0$ as required.
\end{proof}

Note that we have here also a fixed decoder; if we are only
interested in a fixed encoder we can adjust $\alpha_S$ at the
decoder and reduce the margin from optimality.

\subsection{Joint Source/Channel Broadcasting}
\label{sub_broadcast}

\begin{figure*}[!t]\centering
\input{joint_broadcast.pstex_t}
\ccaption{A broadcast presentation of the uncertainty problem.}
\label{broadcast_fig}
\end{figure*}

Abandoning the high SNR assumption, we can no longer simultaneously
approach the optimal performance \eqref{SDR_opt} for multiple SNRs.
However, in many cases we can still do better than a
separation-based scheme. In order to demonstrate that, we choose to
alternate our view to a \emph{broadcast} scenario, where the same
source needs to be transmitted to multiple decoders, each one with
different conditions; yet all the decoders share the same channel
interference $\bI$, see \figref{broadcast_fig}. The variation of the
source SI component $\bJ$ between decoders means that the source has
two decompositions: \beq{J12} \bS = \bQ_1 + \bJ_1 = \bQ_2 + \bJ_2 \
\ , \eeq and we define the per-element variances of the unknown
parts as $\sigma_1^2$ and $\sigma_2^2$, respectively. Note that this
variation does not imply any uncertainty from the point of view of
the MLM encoder, as long as $\sigma_1^2=\sigma_2^2$; see
\cite{Wolf04} for a similar observation in the context of source
coding. We denote the signal-to-noise ratios at the decoders as
$\SNR_1\leq \SNR_2$, and find achievable corresponding
signal-to-distortion ratio $\{\SDR_1,\SDR_2\}$ pairs. It will become
evident from the exposition, that this approach is also good for a
continuum of possible SNRs.

We start from the case $\sigma_1^2=\sigma_2^2$, for which we have
the following.

\vspace{5mm} \begin{theorem} \label{thm_finite_SNR} In the broadcast
WZ/DPC channel of \figref{broadcast_fig} with
$\sigma_1^2=\sigma_2^2$, the signal-to-distortions pair
\[\left\{1 +
\frac{\overline\alpha \cdot
\SNR_1}{\alpha_C^2+(1-\alpha_C)^2\SNR_1},1 + \frac{\overline\alpha
\cdot \SNR_2}{\alpha_C^2+(1-\alpha_C)^2\SNR_2}\right\} \ \ ,
\] where
\beq{beta_alpha_C}
\overline\alpha=\alpha_C\left(2-\frac{\SNR_1+1}{\SNR_1}\alpha_C\right)
\ \ , \eeq can be approached for any $0<\alpha_C\leq
\min\left(1,\frac{2\cdot\SNR_1}{1+\SNR_1}\right)$. In addition, if
there is no channel interference ($\bI=\bzero$), then the pair
$\left\{1+\SNR_1,1 + \frac{\SNR_1(1+\SNR_2)}{1+\SNR_1} \right\}$ can
be approached as well. \vspace{3mm}
\end{theorem}

\begin{figure}[t]
 \subfloat[$\SNR_1=2$, $\SNR_2=10$]{
    \epsfig{file=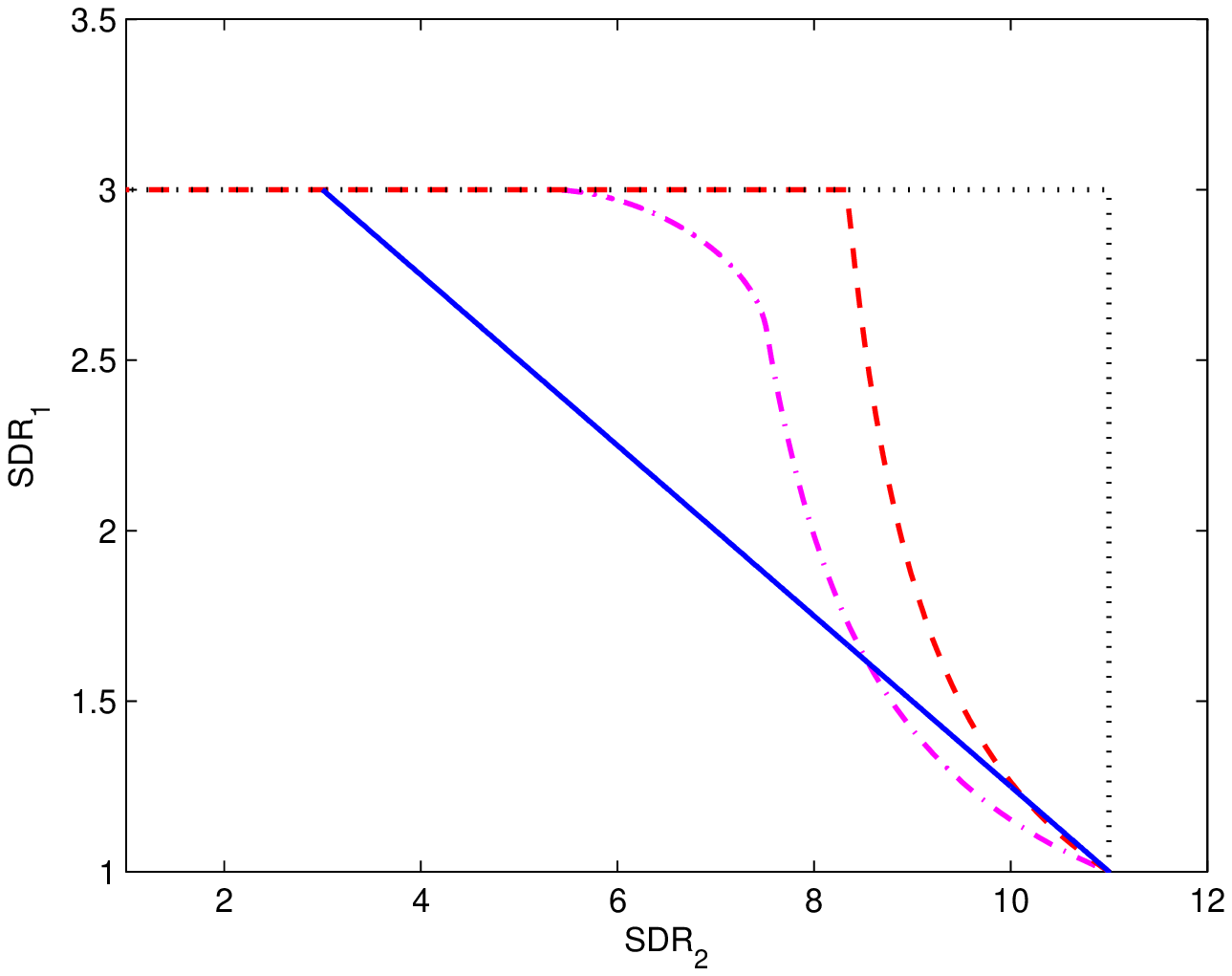, scale = 0.52}}
      \
      \subfloat[$\SNR_1=10$, $\SNR_2=50$]{
    \epsfig{file=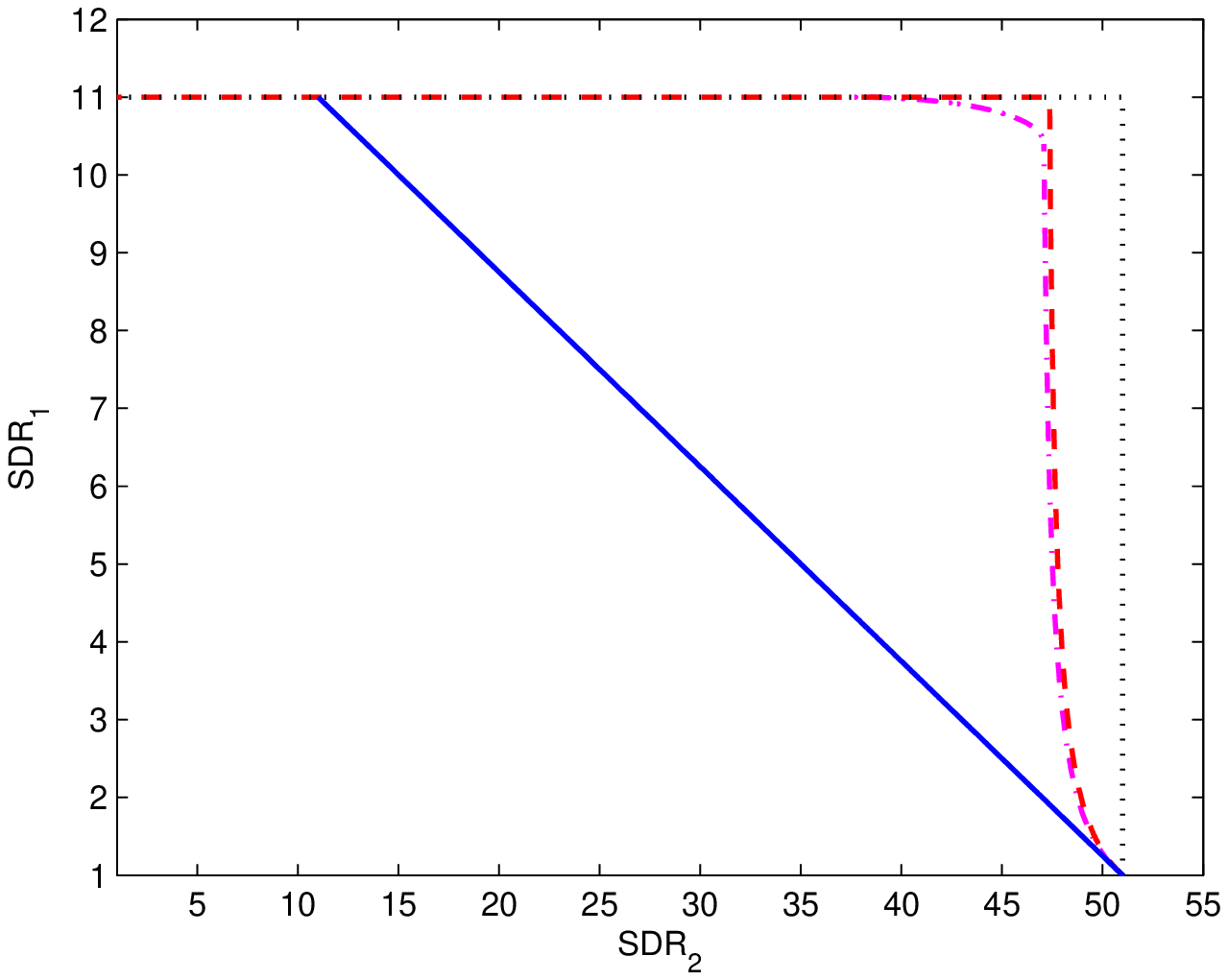, scale = 0.52}}
    \vspace{-5mm} \ccaption{Broadcast performance. Solid line: Achievable by separation for arbitrary $\bI$ and $\bJ$.
    Dash-dotted line: Achievable by MLM for arbitrary $\bI$ and $\bJ$. Dashed line: Achievable by MLM for arbitrary $\bJ$,
    with $\bI=\bzero$. Dotted line:
    Outer bound of ideal matching to both SNRs (achievable by analog transmission when $\bI=\bJ=\bzero$).}
\label{fig_SDRs}
\end{figure}

\begin{proof}
As in the proof of \thref{thm_robust_WZ}, we use
Lemma~\ref{lemma_mismatch} with a choice of $\beta$ which allows
correct decoding in the lower SNR. For the first part of the
theorem, fix any $\alpha_C$ according to the theorem conditions, and
choose any
\[\beta^2< \overline\alpha\frac{P}{\sigma_Q^2} \ \
, \] where $\overline\alpha$ was defined in \eqref{beta_alpha_C}, in
order to satisfy \eqref{mismatch_condition}. In each decoder,
optimize $\alpha_S$ in \eqref{mismatch_SDR} to approach the desired
distortion. For the second part of the theorem, if there is no
channel interference, the encoder is $\alpha_C$-independent, thus
each decoder may work with a different $\alpha_C$ value. We can
therefore make the encoder and the first decoder optimal for
$\SNR_1$, while the second decoder only suffers from the choice of
$\beta$ at the encoder. Again we substitute in \eqref{mismatch_SDR}
to arrive at the desired result
\end{proof}

By standard time-sharing arguments, the achievable SDR regions
include the convex hull (in the distortions plane) defined by these
points and the trivial $\{1+\SNR_1,1\}$ and $\{1,1+\SNR_2\}$ points.
\figref{fig_SDRs} demonstrates these regions, compared to the ideal
(unachievable) region of simultaneous optimality for both SNRs, and
the separation-based region achieved by the concatenation of
successive-refinement source code (see e.g. \cite{EquitzCover91})
with broadcast channel code \cite{CoverBroadcast} (about the
sub-optimality of this combination without SI, see e.g.
\cite{ChenWornell98}). It is evident, that in most cases the use of
the MLM scheme significantly improves the SDR tradeoff over the
performance offered by the separation principle, and that the scheme
approaches simultaneous optimality where both SNRs are high, as
promised by \thref{thm_robust_WZ}. Note that, unlike the
separation-based approach, the MLM approach also offers reasonable
SDRs for intermediate SNRs. Moreover, note that this region is
achievable when no assumption is made about the statistics of $\bI$
and $\bJ$. If these interferences are not very strong comparing to
$P$ and $\sigma_Q^2$, respectively, then one may further extend the
achievable region by allowing some residual interference.

To conclude, we briefly discuss the case where
$\sigma_1^2\neq\sigma_2^2$. We define the SDR of each decoder
relative to its own variance, and ask what are the achievable SDRs
for a pair of SNRs, which may be equal or different. Assume here the
simple case, where there is no channel interference, i.e.
$\bI=\bzero$. In this case, the encoder only needs to agree upon
$\beta$ with the decoders, thus (by Lemma~\ref{lemma_mismatch}) we
may approach for $n=1,2$: \beq{beta_not_opt} \SDR_n = 1 +
\frac{\beta^2}{\beta_{opt,n}^2} \SNR_n \ \ , \eeq where
$\beta_{opt,n}$ is the optimum choice of $\beta$ for $\SNR_n$
according to \eqref{Limit_parameters}. It follows, that if the two
decoders require the same value of $\beta$, they may be both
approach the theoretically optimal distortion. This translates to
the optimality condition:
\[ \sigma_1^2 \frac{1+\SNR_1}{\SNR_1} = \sigma_2^2
\frac{1+\SNR_2}{\SNR_2} \ \ . \] This scenario was presented in
\cite{GunduzNayakTuncelISIT08}, where simultaneous optimality using
hybrid digital/analog schemes was proven under a different
condition:
\[ \frac{\sigma_1^2}{\SNR_1} = \frac{\sigma_2^2}{\SNR_2} \ \ .
\] Both conditions reflect the fact that better source conditions
(lower $\sigma_Q^2$) can compensate for worse channel conditions
(lower SNR). It follows from the difference between the conditions,
that for some parameter values the MLM scheme outperforms the
approach of \cite{GunduzNayakTuncelISIT08}, thus extending the
achievable SDRs region.

\section{Discussion: Delay and Complexity} \label{conclusion}

We have presented the joint source/channel MLM scheme, proven its
optimality for joint WZ/DPC setting with known SNR and shown its
improved robustness over a separation-based scheme. We now discuss
the potential complexity and delay advantages of our approach
relative to separation-based schemes, first considering the
complexity at high dimension and then suggesting a scalar variant.

Consider a separation-based solution, with source and channel
encoder/decoder pairs. An optimal channel coding scheme typically
consists of two codes: an information-bearing code and a shaping
code, both of which require a nearest-neighbor search at the
decoder. An optimal source coding scheme also consists of both a
quantization code and a shaping code in order to achieve the full
vector quantization gain (see e.g. \cite{Gray_al89a}), thus two
nearest-neighbor searches are needed at the encoder. The MLM
approach omits the information-bearing channel code and the
quantization code, and merges the channel and source shaping codes
into one. It is convenient to compare this approach with the nested
lattices approach to channel and source coding with SI
\cite{RamiShamaiUriLattices}, since in that approach both the
channel and source information bearing/shaping code pairs are
materialized by nested lattices. In comparison, our scheme require
only a single lattice (parallel to the coarse lattice of nested
schemes), and in addition the source and channel lattices collapse
into a single one.

There is a price to pay, however: For the WZ problem, the coarse
lattice should be good for channel coding, while for the WDP problem
the coarse lattice should be good for source coding
\cite{RamiShamaiUriLattices}. The lattice used for MLM needs to be
simultaneously good for source \emph{and} channel coding (see
Appendix~\ref{appendix_lattice}). While the existence of such
lattices in the high dimension limit is assured by
\cite{GoodLattices}, in finite dimension the lattice that is best in
one sense is not necessarily best in the other sense
\cite{Conway88}, resulting in a larger implementation loss.
Quantitively, whereas for source coding the lattice should have a
low normalized second moment, and for channel coding it should have
a low volume-to-noise ratio, for joint source channel coding the
\emph{product} $L(\Lambda,p_e)$ \eqref{simple_L_def} should be
low\footnote{In \thref{thm_any_lattice} we show that the figure of
merit is $L(\Lambda,p_e,\alpha)$ \eqref{L_def}, but for reasonably
high SNR it seems that the effect of self noise should not be too
dominant, so we can set $\alpha=1$.} (see
Appendix~\ref{appendix_lattice}). The study of such lattices is
currently under research. Exact comparison of schemes in high
dimension will involve studying the achieved \emph{joint
source/channel excess distortion exponent} (see
\cite{ZhongGaussianJointExponent} for a recent work about this
exponent in the Gaussian setting).

\begin{figure*}[!t]\begin{center}
\input{joint_scheme_companding.pstex_t} \end{center}
\ccaption{Scalar MLM/companding scheme for joint source/channel
coding over a high-SNR dirty-paper channel: $S$ = source, $\hat S$ =
reconstruction, $Z$ = channel noise, $I$ = interference known at the
encoder, $g(\cdot)$~=~companding function.} \label{companding_fig}
\end{figure*}

From the practical point of view, the question of a low-dimensional
scheme is very important, since it implies both low complexity and
low delay. One may ask then, what can be achieved using
low-dimensional lattices, e.g. a scalar lattice? The difficulty,
however, is that in low dimensions a low probability of incorrect
decoding $p_e$ implies a high loss factor $L(\Lambda,p_e)$, thus the
distortion promised by \thref{thm_any_lattice} grows. Some
improvement may be achieved by using an optimal decoder rather than
the one described in this work (see Remark 1 at the end of
\secref{Sec_Joint}), an issue which is left for further research. A
recent work \cite{ItaiThesis} suggests an alternative, for the case
of channel interference only ($\bJ=\bzero$), by also changing the
encoder: The scalar zooming factor $\beta$ of the MLM scheme is
replaced by non-linear companding of the signal; see
\figref{companding_fig}. At high SNR, the distortion loss of such a
scalar MLM scheme with optimal companding comparing to \eqref{OPTA}
is shown to be
\[\frac{D^{companding}}{D^{opt}} = \frac{\sqrt{3}\pi}{2} \cong 4.3dB \ \ . \]
In comparison, the loss of a separation-based scalar scheme,
consisting of a scalar quantizer and a scalar (uncoded) channel
constellation, is \emph{unbounded} in the limit
$\SNR\rightarrow\infty$. This is since in a separation-based scheme
the mapping of quantized source values to channel inputs is
arbitrary; consequently, keeping the loss bounded implies that the
error probability must go to zero in the high-SNR limit, and the gap
of a scalar constellation from capacity grows.

\section*{Acknowledgement}

We thank Uri Erez for helping to make some of the connections which
led to this work.

\appendices
\section{Measures of Goodness of Lattices}
\label{appendix_lattice}

In this appendix we discuss measures of goodness of lattices for
source and channel coding, and their connection with the loss factor
relevant to our joint source/channel scheme.

When a lattice is used as a quantization codebook in the quadratic
Gaussian setting, the figure of merit is the lattice
\emph{normalized second moment}: \beq{G} G(\Lambda) \Ddef
\frac{\sigma^2(\Lambda)}{V(\Lambda)^{\frac{2}{K}}} \ \ , \eeq where
the cell volume is $V(\Lambda)=\int_{\Nu_0} d\bx$. By the
isoperimetric inequality, $G(\Lambda) \geq G^*_K$, where $G^*_K$ is
the normalized second moment of a ball with the same dimension $K$
as the lattice. This quantity satisfies $G^*_K \geq \frac{1}{2\pi
e}$, with asymptotic equality in the limit of large dimension. A
sequence of $K$-dimensional lattices is said to be \emph{good for
MSE quantization} if \beq{good1}\lim_{K\rightarrow\infty}
G(\Lambda_K) = \frac{1}{2\pi e} \ \ , \eeq thus it asymptotically
achieves the minimum possible lattice second moment for a given
volume.

When a lattice is used as an AWGN channel codebook, the figure of
merit is the lattice \emph{volume-to-noise ratio} at a given error
probability $1>p_e>0$ (see e.g. \cite{Forney00,UriRamiAWGN}):
\beq{VNR} \mu(\Lambda,p_e) \Ddef
\frac{V(\Lambda)^{\frac{2}{K}}}{\sigma_Z^2} \ \ , \eeq where
$\sigma_Z^2$ is the maximum variance (per element) of a white
Gaussian vector $\bZ$ having an error probability \[\Pr\{\bZ \notin
\Nu_0 \} \leq p_e \ \ . \] For any lattice, $\mu(\Lambda,p_e) \geq
\mu^*_K(p_e)$, where $\mu^*_K(p_e)$ is the volume-to-noise ratio of
a ball with the same dimension $K$ as the lattice. For any
$1>p_e>0$, $\mu^*_K(p_e) \geq 2\pi e$, with asymptotic equality in
the limit of large dimension. A sequence of $K$-dimensional lattices
is \emph{good for AWGN channel coding} if \beq{good2}
\lim_{p_e\rightarrow 0} \lim_{K\rightarrow\infty} \mu(\Lambda_K,p_e)
= 2\pi e \ \ , \eeq thus it possesses the property of having a
minimum possible cell volume such that the probability of an i.i.d.
Gaussian vector of a given power to fall outside the cell vanishes.

Combining the definitions \eqref{G} and \eqref{VNR}, we see that the
loss factor $L(\Lambda,p_e)$ \eqref{simple_L_def} satisfies:
\[L(\Lambda,p_e) = G(\Lambda) \cdot \mu(\Lambda,p_e) \ \ . \]
Furthermore, the existence of a good sequence of lattices in the
sense of \eqref{good_L} is assured by the existence of a sequence
that simultaneously satisfies \eqref{good1} \emph{and}
\eqref{good2}, which was shown in \cite[Theorem 5]{GoodLattices}.

Proposition~\ref{prop_lattice} is implicit in the proof of
\cite[Theorem 5]{UriRamiAWGN}. It is based upon the existence of
lattices that are simultaneously good for AWGN channel coding and
for covering \cite{GoodLattices}, where goodness for covering also
implies goodness for MSE quantization; for such lattices, it is
shown that the mixture noise cannot be much worse than a Gaussian
noise of the same variance. Later, it was shown in
\cite{LiuMulinKoetter06} that, for such lattices, for small enough
error probability $p_e$, the introduction of self noise actually
reduces the loss factor, i.e. $L(\Lambda,p_e,\alpha)\leq
L(\Lambda,p_e,1)$.

\section{The effect of Decoding Failure on the Distortion}

With probability $p_e$, correct lattice decoding fails, i.e.
\eqref{in_cell} does not hold. These events contribute to the total
distortion a portion of \beq{tilde_D} \tilde D \Ddef p_e \cdot
D^{incorrect} \ \ , \eeq where $D^{incorrect}$ is the distortion
given a decoding failure, as in the proof of
\thref{thm_any_lattice}. In this Appendix we quantify this effect:
In the first part we show that $D^{max}$ of \eqref{d_max} is a
(rather loose) bound on $D^{incorrect}$, thus completing the proof
of \thref{thm_any_lattice}. In the second part, we show directly
that $\tilde D$ must vanish in the limit of small $p_e$, without
resorting to an explicit bound on $D^{incorrect}$.

In both parts we use the observation that \beq{hat_Q} \hat \bS -
\bS\ = \hat \bQ - \bQ \ \ , \eeq where $\hat
\bQ\Ddef\frac{\alpha_S}{\beta}[\beta\bQ+\bZ_{eq}]\bmod\Lambda$, see
also \figref{output_eq_fig}. We note that although $\bQ$ is
unbounded, we always have that \beq{bounded_hat_Q} \hat \bQ
\in\frac{\alpha_S}{\beta}\Nu_0 \ \ . \eeq

\subsection{A Bound on the Conditional Distortion for Any Lattice}
\label{appendix_D2}

In order to complete the proof of \thref{thm_any_lattice}, we now
bound $D^{incorrect}$ of \eqref{D12}.
 \beqn{D2} D^{incorrect} &=& \frac{1}{K} E\{\| \hat
\bS - \bS\|^2\ | \beta \bQ + \bZ_{eq} \notin \Nu_0\} \nonumber \\
&=& \frac{1}{K} E\{\| \hat\bQ - \bQ\|^2 |
\beta \bQ + \bZ_{eq} \notin \Nu_0\} \nonumber \\
&\leq& \frac{2}{K} \Bigl(E\{\| \hat\bQ \|^2 | \beta \bQ + \bZ_{eq}
\notin \Nu_0\}  + E\{\| \bQ\|^2 | \beta \bQ + \bZ_{eq} \notin \Nu_0
\}\Bigr) \ \ , \eeqn where the inequality follows from assuming
maximizing $(-1)$ correlation coefficient and then applying the
Cauchy-Schwartz inequality. We shall now bound these two terms. For
the first one, recalling the definition of the covering radius
\eqref{covering_radius}, we bound the conditional expectation by the
maximum possible value: \beq{D_2_1} E\{\| \hat\bQ \|^2 | \beta \bQ +
\bZ_{eq} \notin \Nu_0\} \leq \max (\| \hat\bQ \|^2) =
\frac{\alpha_S^2 \cdot r^2(\Lambda)}{\beta^2} \leq
\frac{r^2(\Lambda)}{\beta^2} \ \ . \eeq For the second term, we
have: \beqn{D_2_2} E\{\| \bQ\|^2 | \beta \bQ + \bZ_{eq} \notin \Nu_0
\} &\leq& E\{\| \bQ\|^2 | \beta \bQ \notin \Nu_0 \} \nonumber
\\ &\leq& E\{\| \bQ\|^2 | \beta \bQ \notin \cB_0 \} \ \ , \nonumber
\eeqn where $\cB_0$ is the circumsphere of $\Nu_0$, of radius
$r(\Lambda)$. It follows that \[ E\{\| \bQ\|^2 | \beta \bQ +
\bZ_{eq} \notin \Nu_0 \} \leq \sigma_Q^2 E \{ V | V > v_0 \} \ \ ,
\] where $V \sim \cX_K^2$ and $v_0 \Ddef \frac{r^2(\Lambda)}{\beta^2
\sigma_Q^2}$. This conditional expectation is given by: \[ E \{ V |
V > v_0\} =
\frac{\cQ(\frac{K}{2}+1,v_0\frac{K}{2})}{\cQ(\frac{K}{2},v_0\frac{K}{2})}
\leq v_0+2 \ \ , \] where $\cQ(\cdot,\cdot)$ is the regularized
incomplete Gamma function, and the inequality can be shown by means
of calculus. This gives the bound on the second term:
\[E\{\| \bQ\|^2 | \beta
\bQ + \bZ_{eq} \notin \Nu_0 \} \leq
\left(\frac{r^2(\Lambda)}{\beta^2} + 2K\sigma_Q^2 \right) \ \ . \]
Substituting this and \eqref{D_2_1} in \eqref{D2}, we have that:
\[ D^{incorrect} \leq 4 \left(\frac{r^2(\Lambda)}{K\beta^2} + \sigma_Q^2 \right) \ \
.
\] Recalling the choice of $\beta$ in \eqref{beta} and the definition of $\tilde
L(\cdot,\cdot)$ in \eqref{tilde_L_def}, the bound follows.

\subsection{Asymptotic Effect of Decoding Failures}
\label{Appendix_Wyner}

In this part we follow the claims used by Wyner in the source coding
context to establish \cite[(5.2)]{Wyner78}, to see that
$\lim_{p_e\rightarrow 0} \tilde D = 0$, where $\tilde D$ was defined
in \eqref{tilde_D}, without using the explicit bound derived in
Appendix~\ref{appendix_D2}. This serves as a simpler proof of
\thref{thm_joint}; moreover, it also applies to a non-optimal choice
of parameters, thus it serves in the analysis of performance under
uncertainty conditions.

Denoting the decoding failure event by $\varepsilon$ and its
indicator by $I_\varepsilon$, and recalling \eqref{hat_Q}, we
re-write the contribution to the distortion as:
\[\tilde D = E\{ I_\varepsilon\cdot (\hat \bQ - \bQ)^2 \} \ \ . \]
For any value of the source unknown part $\bQ$, the distortion is
bounded by:
\[d(\bQ) \Ddef \sup_{\hat\bQ} (\hat \bQ - \bQ)^2 \ \ . \] The
expectation $E\{d(\bQ)\}$ is finite, since $\bQ$ is Gaussian and
$\hat \bQ$ is bounded (see \eqref{bounded_hat_Q}). We now have that
\[\tilde D \leq E\{ I_\varepsilon\cdot d(\bQ) \} \ \ . \] Using a simple
lemma of Probability Theory \cite[Lemma 5.1]{Wyner78}, since
$E\{d(\bQ)\}$ is finite, this expectation approaches zero as
$p(\varepsilon)=p_e\rightarrow 0$.

\bibliographystyle{../latex/IEEEtran}
\bibliography{../latex/mybib}

\end{document}

%% file: joint_prob2.pstex_t
\begin{picture}(0,0)%
\includegraphics{joint_prob2.pstex}%
\end{picture}%
\setlength{\unitlength}{2171sp}%
\begingroup\makeatletter\ifx\SetFigFont\undefined%
\gdef\SetFigFont#1#2#3#4#5{%
  \reset@font\fontsize{#1}{#2pt}%
  \fontfamily{#3}\fontseries{#4}\fontshape{#5}%
  \selectfont}%
\fi\endgroup%
\begin{picture}(9059,2390)(6,-1603)
\put(5101,314){\makebox(0,0)[lb]{\smash{{\SetFigFont{9}{10.8}{\rmdefault}{\mddefault}{\updefault}$\bI$}}}}
\put(7758,-211){\makebox(0,0)[lb]{\smash{{\SetFigFont{8}{9.6}{\rmdefault}{\mddefault}{\updefault}$\hat \bS$}}}}
\put(2552,437){\makebox(0,0)[lb]{\smash{{\SetFigFont{8}{9.6}{\rmdefault}{\mddefault}{\updefault}side information}}}}
\put(2942,637){\makebox(0,0)[lb]{\smash{{\SetFigFont{8}{9.6}{\rmdefault}{\mddefault}{\updefault}Channel}}}}
\put(2087,-1603){\makebox(0,0)[lb]{\smash{{\SetFigFont{8}{9.6}{\rmdefault}{\mddefault}{\updefault}Source side information}}}}
\put(867,-421){\makebox(0,0)[lb]{\smash{{\SetFigFont{9}{10.8}{\rmdefault}{\mddefault}{\updefault}$\Sigma$}}}}
\put(4876,-376){\makebox(0,0)[lb]{\smash{{\SetFigFont{9}{10.8}{\rmdefault}{\mddefault}{\updefault}$\Sigma$}}}}
\put(2606,-391){\makebox(0,0)[lb]{\smash{{\SetFigFont{7}{8.4}{\rmdefault}{\mddefault}{\updefault}ENCODER}}}}
\put(6442,-365){\makebox(0,0)[lb]{\smash{{\SetFigFont{7}{8.4}{\rmdefault}{\mddefault}{\updefault}DECODER}}}}
\put(1686,-216){\makebox(0,0)[lb]{\smash{{\SetFigFont{8}{9.6}{\rmdefault}{\mddefault}{\updefault}$\bS$}}}}
\put(1412,-563){\makebox(0,0)[lb]{\smash{{\SetFigFont{8}{9.6}{\rmdefault}{\mddefault}{\updefault}Source}}}}
\put(192,-139){\makebox(0,0)[lb]{\smash{{\SetFigFont{8}{9.6}{\rmdefault}{\mddefault}{\updefault}$\bQ$}}}}
\put(181,-1261){\makebox(0,0)[lb]{\smash{{\SetFigFont{9}{10.8}{\rmdefault}{\mddefault}{\updefault}$\bJ$}}}}
\put(4452,-1036){\makebox(0,0)[lb]{\smash{{\SetFigFont{7}{8.4}{\rmdefault}{\mddefault}{\updefault}CHANNEL}}}}
\put(3916,-193){\makebox(0,0)[lb]{\smash{{\SetFigFont{8}{9.6}{\rmdefault}{\mddefault}{\updefault}$\bX$}}}}
\put(5839,-179){\makebox(0,0)[lb]{\smash{{\SetFigFont{8}{9.6}{\rmdefault}{\mddefault}{\updefault}$\bY$}}}}
\put(4681,-811){\makebox(0,0)[lb]{\smash{{\SetFigFont{8}{9.6}{\rmdefault}{\mddefault}{\updefault}$\bZ$}}}}
\put(7727,-548){\makebox(0,0)[lb]{\smash{{\SetFigFont{8}{9.6}{\rmdefault}{\mddefault}{\updefault}Reconstruction}}}}
\end{picture}%

%% file: joint_scheme4.pstex_t
\begin{picture}(0,0)%
\includegraphics{joint_scheme4.pstex}%
\end{picture}%
\setlength{\unitlength}{2368sp}%
\begingroup\makeatletter\ifx\SetFigFont\undefined%
\gdef\SetFigFont#1#2#3#4#5{%
  \reset@font\fontsize{#1}{#2pt}%
  \fontfamily{#3}\fontseries{#4}\fontshape{#5}%
  \selectfont}%
\fi\endgroup%
\begin{picture}(11206,1830)(179,-1425)
\put(10848,-136){\makebox(0,0)[lb]{\smash{{\SetFigFont{8}{9.6}{\rmdefault}{\mddefault}{\updefault}$\hat \bS$}}}}
\put(10248,-436){\makebox(0,0)[lb]{\smash{{\SetFigFont{10}{12.0}{\rmdefault}{\mddefault}{\updefault}$\Sigma$}}}}
\put(10473,-1036){\makebox(0,0)[lb]{\smash{{\SetFigFont{8}{9.6}{\rmdefault}{\mddefault}{\updefault}$\bJ$}}}}
\put(9273,-436){\makebox(0,0)[lb]{\smash{{\SetFigFont{8}{9.6}{\rmdefault}{\mddefault}{\updefault}$\frac{\alpha_S}{\beta}$}}}}
\put(1801,-436){\makebox(0,0)[lb]{\smash{{\SetFigFont{7}{8.4}{\rmdefault}{\mddefault}{\updefault}$\beta$}}}}
\put(751,-436){\makebox(0,0)[lb]{\smash{{\SetFigFont{10}{12.0}{\rmdefault}{\mddefault}{\updefault}$\Sigma$}}}}
\put(2626,-436){\makebox(0,0)[lb]{\smash{{\SetFigFont{10}{12.0}{\rmdefault}{\mddefault}{\updefault}$\Sigma$}}}}
\put(2476,-811){\makebox(0,0)[lb]{\smash{{\SetFigFont{8}{9.6}{\rmdefault}{\mddefault}{\updefault}$-$}}}}
\put(6826,-1111){\makebox(0,0)[lb]{\smash{{\SetFigFont{8}{9.6}{\rmdefault}{\mddefault}{\updefault}$\beta \bJ$}}}}
\put(6376,-811){\makebox(0,0)[lb]{\smash{{\SetFigFont{8}{9.6}{\rmdefault}{\mddefault}{\updefault}$-$}}}}
\put(6301,-61){\makebox(0,0)[lb]{\smash{{\SetFigFont{8}{9.6}{\rmdefault}{\mddefault}{\updefault}$-$}}}}
\put(5251,-136){\makebox(0,0)[lb]{\smash{{\SetFigFont{8}{9.6}{\rmdefault}{\mddefault}{\updefault}$\bY$}}}}
\put(3376,-436){\makebox(0,0)[lb]{\smash{{\SetFigFont{8}{9.6}{\rmdefault}{\mddefault}{\updefault}mod $\Lambda$}}}}
\put(5776,-436){\makebox(0,0)[lb]{\smash{{\SetFigFont{8}{9.6}{\rmdefault}{\mddefault}{\updefault}$\alpha_C$}}}}
\put(4576,-136){\makebox(0,0)[lb]{\smash{{\SetFigFont{8}{9.6}{\rmdefault}{\mddefault}{\updefault}$\bX$}}}}
\put(2851,-1111){\makebox(0,0)[lb]{\smash{{\SetFigFont{8}{9.6}{\rmdefault}{\mddefault}{\updefault}$\alpha \bI$}}}}
\put(376,-1411){\makebox(0,0)[lb]{\smash{{\SetFigFont{8}{9.6}{\rmdefault}{\mddefault}{\updefault}SOURCE}}}}
\put(4501,-1411){\makebox(0,0)[lb]{\smash{{\SetFigFont{8}{9.6}{\rmdefault}{\mddefault}{\updefault}CHANNEL}}}}
\put(5070,204){\makebox(0,0)[lb]{\smash{{\SetFigFont{8}{9.6}{\rmdefault}{\mddefault}{\updefault}$\bZ$}}}}
\put(5083,-1020){\makebox(0,0)[lb]{\smash{{\SetFigFont{8}{9.6}{\rmdefault}{\mddefault}{\updefault}$\bI$}}}}
\put(2155,-1411){\makebox(0,0)[lb]{\smash{{\SetFigFont{8}{9.6}{\rmdefault}{\mddefault}{\updefault}ENCODER}}}}
\put(4843,-436){\makebox(0,0)[lb]{\smash{{\SetFigFont{10}{12.0}{\rmdefault}{\mddefault}{\updefault}$\Sigma$}}}}
\put(6568,-436){\makebox(0,0)[lb]{\smash{{\SetFigFont{10}{12.0}{\rmdefault}{\mddefault}{\updefault}$\Sigma$}}}}
\put(213,-205){\makebox(0,0)[lb]{\smash{{\SetFigFont{8}{9.6}{\rmdefault}{\mddefault}{\updefault}$\bQ$}}}}
\put(1284,-211){\makebox(0,0)[lb]{\smash{{\SetFigFont{8}{9.6}{\rmdefault}{\mddefault}{\updefault}$\bS$}}}}
\put(976,-1014){\makebox(0,0)[lb]{\smash{{\SetFigFont{10}{12.0}{\rmdefault}{\mddefault}{\updefault}$\bJ$}}}}
\put(6826,200){\makebox(0,0)[lb]{\smash{{\SetFigFont{8}{9.6}{\rmdefault}{\mddefault}{\updefault}$\bD$}}}}
\put(2826,200){\makebox(0,0)[lb]{\smash{{\SetFigFont{8}{9.6}{\rmdefault}{\mddefault}{\updefault}$\bD$}}}}
\put(6979,-136){\makebox(0,0)[lb]{\smash{{\SetFigFont{8}{9.6}{\rmdefault}{\mddefault}{\updefault}$\bT$}}}}
\put(7623,-436){\makebox(0,0)[lb]{\smash{{\SetFigFont{8}{9.6}{\rmdefault}{\mddefault}{\updefault}mod $\Lambda$}}}}
\put(7573,-1425){\makebox(0,0)[lb]{\smash{{\SetFigFont{8}{9.6}{\rmdefault}{\mddefault}{\updefault}DECODER}}}}
\put(8701,-211){\makebox(0,0)[lb]{\smash{{\SetFigFont{8}{9.6}{\rmdefault}{\mddefault}{\updefault}$\bM$}}}}
\end{picture}%

%% file: finite_k_equivalent2.pstex_t
\begin{picture}(0,0)%
\includegraphics{finite_k_equivalent2.pstex}%
\end{picture}%
\setlength{\unitlength}{2013sp}%
\begingroup\makeatletter\ifx\SetFigFont\undefined%
\gdef\SetFigFont#1#2#3#4#5{%
  \reset@font\fontsize{#1}{#2pt}%
  \fontfamily{#3}\fontseries{#4}\fontshape{#5}%
  \selectfont}%
\fi\endgroup%
\begin{picture}(7824,1674)(4789,-2323)
\put(10201,-1186){\makebox(0,0)[lb]{\smash{{\SetFigFont{8}{9.6}{\rmdefault}{\mddefault}{\updefault}$\frac{\alpha_S}{\beta}$}}}}
\put(10951,-1411){\makebox(0,0)[lb]{\smash{{\SetFigFont{8}{9.6}{\rmdefault}{\mddefault}{\updefault}$\hat \bQ$}}}}
\put(11551,-1261){\makebox(0,0)[lb]{\smash{{\SetFigFont{9}{10.8}{\rmdefault}{\mddefault}{\updefault}$\Sigma$}}}}
\put(11326,-2086){\makebox(0,0)[lb]{\smash{{\SetFigFont{8}{9.6}{\rmdefault}{\mddefault}{\updefault}$\bJ$}}}}
\put(12151,-1411){\makebox(0,0)[lb]{\smash{{\SetFigFont{8}{9.6}{\rmdefault}{\mddefault}{\updefault}$\hat \bS$}}}}
\put(11401,-1636){\makebox(0,0)[lb]{\smash{{\SetFigFont{6}{7.2}{\rmdefault}{\mddefault}{\updefault}$+$}}}}
\put(7126,-1261){\makebox(0,0)[lb]{\smash{{\SetFigFont{9}{10.8}{\rmdefault}{\mddefault}{\updefault}$\Sigma$}}}}
\put(5926,-1186){\makebox(0,0)[lb]{\smash{{\SetFigFont{8}{9.6}{\rmdefault}{\mddefault}{\updefault}$\beta$}}}}
\put(5101,-1411){\makebox(0,0)[lb]{\smash{{\SetFigFont{8}{9.6}{\rmdefault}{\mddefault}{\updefault}$\bQ$}}}}
\put(6901,-1561){\makebox(0,0)[lb]{\smash{{\SetFigFont{6}{7.2}{\rmdefault}{\mddefault}{\updefault}$+$}}}}
\put(6751,-2086){\makebox(0,0)[lb]{\smash{{\SetFigFont{8}{9.6}{\rmdefault}{\mddefault}{\updefault}$\bZ_{eq}$}}}}
\put(8401,-1186){\makebox(0,0)[lb]{\smash{{\SetFigFont{8}{9.6}{\rmdefault}{\mddefault}{\updefault}$\mod \Lambda$}}}}
\put(7651,-1411){\makebox(0,0)[lb]{\smash{{\SetFigFont{8}{9.6}{\rmdefault}{\mddefault}{\updefault}$\bT$}}}}
\put(9601,-1411){\makebox(0,0)[lb]{\smash{{\SetFigFont{8}{9.6}{\rmdefault}{\mddefault}{\updefault}$\bM$}}}}
\end{picture}%

%% file: output_equivalent4.pstex_t
\begin{picture}(0,0)%
\includegraphics{output_equivalent4.pstex}%
\end{picture}%
\setlength{\unitlength}{2013sp}%
\begingroup\makeatletter\ifx\SetFigFont\undefined%
\gdef\SetFigFont#1#2#3#4#5{%
  \reset@font\fontsize{#1}{#2pt}%
  \fontfamily{#3}\fontseries{#4}\fontshape{#5}%
  \selectfont}%
\fi\endgroup%
\begin{picture}(6222,2222)(4789,-2323)
\put(5101,-1411){\makebox(0,0)[lb]{\smash{{\SetFigFont{8}{9.6}{\rmdefault}{\mddefault}{\updefault}$\bQ$}}}}
\put(6901,-1561){\makebox(0,0)[lb]{\smash{{\SetFigFont{6}{7.2}{\rmdefault}{\mddefault}{\updefault}$+$}}}}
\put(6751,-2086){\makebox(0,0)[lb]{\smash{{\SetFigFont{8}{9.6}{\rmdefault}{\mddefault}{\updefault}$\bZ_{eq}$}}}}
\put(9349,-1411){\makebox(0,0)[lb]{\smash{{\SetFigFont{8}{9.6}{\rmdefault}{\mddefault}{\updefault}$\hat \bQ$}}}}
\put(10549,-1411){\makebox(0,0)[lb]{\smash{{\SetFigFont{8}{9.6}{\rmdefault}{\mddefault}{\updefault}$\hat \bS$}}}}
\put(9799,-1636){\makebox(0,0)[lb]{\smash{{\SetFigFont{6}{7.2}{\rmdefault}{\mddefault}{\updefault}$+$}}}}
\put(9824,-2086){\makebox(0,0)[lb]{\smash{{\SetFigFont{8}{9.6}{\rmdefault}{\mddefault}{\updefault}$\bJ$}}}}
\put(6056,-1186){\makebox(0,0)[lb]{\smash{{\SetFigFont{8}{9.6}{\rmdefault}{\mddefault}{\updefault}$\beta$}}}}
\put(7266,-1181){\makebox(0,0)[lb]{\smash{{\SetFigFont{9}{10.8}{\rmdefault}{\mddefault}{\updefault}$\Sigma$}}}}
\put(8691,-1156){\makebox(0,0)[lb]{\smash{{\SetFigFont{8}{9.6}{\rmdefault}{\mddefault}{\updefault}$\frac{\alpha_S}{\beta}$}}}}
\put(10082,-1181){\makebox(0,0)[lb]{\smash{{\SetFigFont{9}{10.8}{\rmdefault}{\mddefault}{\updefault}$\Sigma$}}}}
\put(8763,-257){\makebox(0,0)[lb]{\smash{{\SetFigFont{6}{7.2}{\rmdefault}{\mddefault}{\updefault}Power Constraint $\frac{P}{L(\Lambda,p_e,\alpha_C)}$}}}}
\put(7651,-1411){\makebox(0,0)[lb]{\smash{{\SetFigFont{8}{9.6}{\rmdefault}{\mddefault}{\updefault}$\bM$}}}}
\end{picture}%

%% file: joint_broadcast.pstex_t
\begin{picture}(0,0)%
\includegraphics{joint_broadcast.pstex}%
\end{picture}%
\setlength{\unitlength}{2171sp}%
\begingroup\makeatletter\ifx\SetFigFont\undefined%
\gdef\SetFigFont#1#2#3#4#5{%
  \reset@font\fontsize{#1}{#2pt}%
  \fontfamily{#3}\fontseries{#4}\fontshape{#5}%
  \selectfont}%
\fi\endgroup%
\begin{picture}(7749,3024)(1864,-1723)
\put(4876,-376){\makebox(0,0)[lb]{\smash{{\SetFigFont{9}{10.8}{\rmdefault}{\mddefault}{\updefault}$\Sigma$}}}}
\put(6151,299){\makebox(0,0)[lb]{\smash{{\SetFigFont{9}{10.8}{\rmdefault}{\mddefault}{\updefault}$\Sigma$}}}}
\put(6151,-826){\makebox(0,0)[lb]{\smash{{\SetFigFont{9}{10.8}{\rmdefault}{\mddefault}{\updefault}$\Sigma$}}}}
\put(2606,-391){\makebox(0,0)[lb]{\smash{{\SetFigFont{7}{8.4}{\rmdefault}{\mddefault}{\updefault}ENCODER}}}}
\put(7717,-740){\makebox(0,0)[lb]{\smash{{\SetFigFont{7}{8.4}{\rmdefault}{\mddefault}{\updefault}DECODER}}}}
\put(7717,385){\makebox(0,0)[lb]{\smash{{\SetFigFont{7}{8.4}{\rmdefault}{\mddefault}{\updefault}DECODER}}}}
\put(8026, 89){\makebox(0,0)[lb]{\smash{{\SetFigFont{8}{9.6}{\rmdefault}{\mddefault}{\updefault}$1$}}}}
\put(8026,-1036){\makebox(0,0)[lb]{\smash{{\SetFigFont{8}{9.6}{\rmdefault}{\mddefault}{\updefault}$2$}}}}
\put(8326,1064){\makebox(0,0)[lb]{\smash{{\SetFigFont{9}{10.8}{\rmdefault}{\mddefault}{\updefault}$\bJ_1$}}}}
\put(8326,-1636){\makebox(0,0)[lb]{\smash{{\SetFigFont{9}{10.8}{\rmdefault}{\mddefault}{\updefault}$\bJ_2$}}}}
\put(4351,-1111){\makebox(0,0)[lb]{\smash{{\SetFigFont{7}{8.4}{\rmdefault}{\mddefault}{\updefault}BROADCAST}}}}
\put(4501,-1336){\makebox(0,0)[lb]{\smash{{\SetFigFont{7}{8.4}{\rmdefault}{\mddefault}{\updefault}CHANNEL}}}}
\put(3916,-286){\makebox(0,0)[lb]{\smash{{\SetFigFont{8}{9.6}{\rmdefault}{\mddefault}{\updefault}$\bX$}}}}
\put(1876,-286){\makebox(0,0)[lb]{\smash{{\SetFigFont{8}{9.6}{\rmdefault}{\mddefault}{\updefault}$\bS$}}}}
\put(9033,-736){\makebox(0,0)[lb]{\smash{{\SetFigFont{8}{9.6}{\rmdefault}{\mddefault}{\updefault}$\hat \bS_2$}}}}
\put(9033,389){\makebox(0,0)[lb]{\smash{{\SetFigFont{8}{9.6}{\rmdefault}{\mddefault}{\updefault}$\hat \bS_1$}}}}
\put(9033, 89){\makebox(0,0)[lb]{\smash{{\SetFigFont{8}{9.6}{\rmdefault}{\mddefault}{\updefault}$\SDR_1$}}}}
\put(9033,-1036){\makebox(0,0)[lb]{\smash{{\SetFigFont{8}{9.6}{\rmdefault}{\mddefault}{\updefault}$\SDR_2$}}}}
\put(4426,464){\makebox(0,0)[lb]{\smash{{\SetFigFont{9}{10.8}{\rmdefault}{\mddefault}{\updefault}$\bI$}}}}
\put(6001,689){\makebox(0,0)[lb]{\smash{{\SetFigFont{8}{9.6}{\rmdefault}{\mddefault}{\updefault}$\bZ_1 \ \SNR_1$}}}}
\put(7051,389){\makebox(0,0)[lb]{\smash{{\SetFigFont{8}{9.6}{\rmdefault}{\mddefault}{\updefault}$\bY_1$}}}}
\put(7051,-736){\makebox(0,0)[lb]{\smash{{\SetFigFont{8}{9.6}{\rmdefault}{\mddefault}{\updefault}$\bY_2$}}}}
\put(5851,-1336){\makebox(0,0)[lb]{\smash{{\SetFigFont{8}{9.6}{\rmdefault}{\mddefault}{\updefault}$\bZ_2 \ \ \SNR_2$}}}}
\end{picture}%

%% file: joint_scheme_companding.pstex_t
\begin{picture}(0,0)%
\includegraphics{joint_scheme_companding.pstex}%
\end{picture}%
\setlength{\unitlength}{2368sp}%
\begingroup\makeatletter\ifx\SetFigFont\undefined%
\gdef\SetFigFont#1#2#3#4#5{%
  \reset@font\fontsize{#1}{#2pt}%
  \fontfamily{#3}\fontseries{#4}\fontshape{#5}%
  \selectfont}%
\fi\endgroup%
\begin{picture}(9174,1830)(289,-1425)
\put(4501,-1411){\makebox(0,0)[lb]{\smash{{\SetFigFont{8}{9.6}{\rmdefault}{\mddefault}{\updefault}CHANNEL}}}}
\put(5070,204){\makebox(0,0)[lb]{\smash{{\SetFigFont{8}{9.6}{\rmdefault}{\mddefault}{\updefault}$Z$}}}}
\put(2155,-1411){\makebox(0,0)[lb]{\smash{{\SetFigFont{8}{9.6}{\rmdefault}{\mddefault}{\updefault}ENCODER}}}}
\put(4843,-436){\makebox(0,0)[lb]{\smash{{\SetFigFont{10}{12.0}{\rmdefault}{\mddefault}{\updefault}$\Sigma$}}}}
\put(2326,-436){\makebox(0,0)[lb]{\smash{{\SetFigFont{10}{12.0}{\rmdefault}{\mddefault}{\updefault}$\Sigma$}}}}
\put(2176,-811){\makebox(0,0)[lb]{\smash{{\SetFigFont{8}{9.6}{\rmdefault}{\mddefault}{\updefault}$-$}}}}
\put(451,-211){\makebox(0,0)[lb]{\smash{{\SetFigFont{8}{9.6}{\rmdefault}{\mddefault}{\updefault}$S$}}}}
\put(9076,-136){\makebox(0,0)[lb]{\smash{{\SetFigFont{8}{9.6}{\rmdefault}{\mddefault}{\updefault}$\hat S$}}}}
\put(3076,-436){\makebox(0,0)[lb]{\smash{{\SetFigFont{8}{9.6}{\rmdefault}{\mddefault}{\updefault}mod $\Lambda$}}}}
\put(6226,-436){\makebox(0,0)[lb]{\smash{{\SetFigFont{8}{9.6}{\rmdefault}{\mddefault}{\updefault}mod $\Lambda$}}}}
\put(5101,-961){\makebox(0,0)[lb]{\smash{{\SetFigFont{8}{9.6}{\rmdefault}{\mddefault}{\updefault}$I$}}}}
\put(4201,-136){\makebox(0,0)[lb]{\smash{{\SetFigFont{8}{9.6}{\rmdefault}{\mddefault}{\updefault}$X$}}}}
\put(5626,-136){\makebox(0,0)[lb]{\smash{{\SetFigFont{8}{9.6}{\rmdefault}{\mddefault}{\updefault}$Y$}}}}
\put(2551,-961){\makebox(0,0)[lb]{\smash{{\SetFigFont{8}{9.6}{\rmdefault}{\mddefault}{\updefault}$I$}}}}
\put(1201,-586){\makebox(0,0)[lb]{\smash{{\SetFigFont{8}{9.6}{\rmdefault}{\mddefault}{\updefault}$g(\cdot)$}}}}
\put(7876,-586){\makebox(0,0)[lb]{\smash{{\SetFigFont{8}{9.6}{\rmdefault}{\mddefault}{\updefault}$g^{-1}(\cdot)$}}}}
\put(7126,-1425){\makebox(0,0)[lb]{\smash{{\SetFigFont{8}{9.6}{\rmdefault}{\mddefault}{\updefault}DECODER}}}}
\end{picture}%